\documentclass[12pt]{elsarticle}
\renewcommand{\appendixname}{Appendix} % Fixes a mistake in elsarticle when referencing an appendix
\usepackage{microtype}
\usepackage[margin=1in]{geometry}
\usepackage{amsmath,amssymb,amsthm,bm}
\usepackage[capitalise]{cleveref}
\biboptions{sort&compress}
\usepackage{epstopdf}
\usepackage{float}
\usepackage{xcolor}
\usepackage{enumitem}

\usepackage[caption=false]{subfig} % This seems to make the figure syntax much simpler
\usepackage{placeins} %\FloatBarrier

\usepackage{xurl} %\url{}

\AppendGraphicsExtensions{.tif}

\ifdefined\usetodonotes
\fi
\usepackage[colorinlistoftodos,textsize=small]{todonotes}
\definecolor{R1}{HTML}{1F77B4}
\definecolor{C0}{HTML}{1F77B4}
\definecolor{C1}{HTML}{FF7F0E}
\definecolor{C2}{HTML}{2ca02c}
\definecolor{C3}{HTML}{d62728}
\definecolor{C4}{HTML}{9467bd}
\definecolor{C5}{HTML}{8c564b}
\providecommand{\nvec}[1]{\hat{\bm{#1}}}

\begin{document}

\title{Comparison of evolving interfaces, triple points, and quadruple points for discrete and diffuse interface methods}

\author[ucdavis]{Erdem Eren\corref{cor1}}
\author[uccs]{Brandon Runnels}
\author[ucdavis]{Jeremy Mason}

\cortext[cor1]{Corresponding author}

\address[ucdavis]{Department of Materials Science and Engineering, University of California at Davis, Davis, CA, USA}
\address[uccs]{Department of Mechanical and Aerospace Engineering, University of Colorado, Colorado Springs, CO USA}

\begin{abstract}
The evolution of interfaces is intrinsic to many physical processes ranging from cavitation in fluids to recrystallization in solids.
Computational modeling of interface motion entails a number of challenges, many of which are related to the range of topological transitions that can occur over the course of the simulation.
Microstructure evolution in a polycrystalline material that involves grain boundary motion is a particularly complex example due to the extreme variety, heterogeneity, and anisotropy of grain boundary properties.
Accurately modeling this process is essential to determining processing-structure-property relationships in polycrystalline materials though.
Simulations of microstructure evolution in such materials often use diffuse interface methods like the phase field method that are advantageous for their versatility and ease of handling complex geometries but can be prohibitively expensive due to the need for high interface resolution.
Discrete interface methods require fewer grid points and can consequently exhibit better performance but have received comparatively little attention, perhaps due to the difficulties of maintaining the mesh and consistently implementing topological transitions on the grain boundary network.
This work explicitly compares a recently-developed discrete interface method to a multiphase field method on several classical problems relating to microstructure evolution in polcrystalline materials: a shrinking spherical grain, the steady-state triple junction dihedral angle, and the steady-state quadruple point dihedral angle. 
In each case, the discrete method is found to meet or outperform the multiphase field method with respect to accuracy for comparable levels of refinement, demonstrating its potential efficacy as a numerical approach for microstructure evolution in polycrystalline materials.
\end{abstract}


\maketitle

\section{Introduction}
\label{sec:introduction}

The simulation of physical systems often requires the modeling of moving interfaces.
This could involve interfaces at the boundaries between different phases of matter including liquid/gas (e.g.\ cavitation), gas/solid (e.g.\ deflagration, sublimation, deposition), and solid/liquid (e.g.\ melting, solidification), or within a single solid phase.
The grain structure of polycrystalline materials in particular contains an extensive network of interfaces known as grain boundaries that separate grains (contiguous regions with a given crystallographic orientation).
The interfacial dynamics governing this grain boundary network are often complex, leading to topological transitions and rapidly changing grain morphologies that pose a unique type of modeling challenge.

A material's grain structure is essential to its macroscopic properties.
For example, solute segregation to grain boundaries can change the grain boundary cohesive energy \cite{1996InterfaceSciHofmann,gibson2015segregation}, or the material's susceptibility to hydrogen embrittlement \cite{2010JourMechPhysSolNovak,huang2017hydrogen} and stress corrosion cracking \cite{pan1996grain,2004ActaMateSong}.
Grain boundaries can provide preferential sites for the precipitation of a second phase, increasing or decreasing the plasticity of the polycrystal \cite{hong2013effects,singh2014enhancing}.
They provide obstructions to the propagation of slip, with implications for the strength of the material as evidenced by the Hall-Petch equation \cite{1951ProcPhysSocHall,1953JourIronSteelPetch}.
Given such consequences of the grain structure, it is not surprising that a variety of methods to simulate grain boundary motion and the evolution of the grain structure have been proposed in the literature \cite{1986ActaMetalMateSrolovitzGrestAndersonvol1,1989PhilMagBKawasaki,1991PhysRevAHolm,2002AnnuRevMatRsRaabe,1999PhysicaDSteinbach,2012CompPhySaye}.
Many of these represent the grain boundaries implicitly, as the locus of points where an indicator function abruptly changes value; such methods are referred to as diffuse interface methods below.
While this has the advantage of not requiring that changes to the grain boundary topology be handled explicitly, the implicit representation complicates simulations of grain boundaries whose properties depend on their crystallography.
This is not an issue when investigating the properties of generic grain boundary networks since a microstructure with constant and isotropic grain boundary properties is regarded as the canonical model system \cite{mason2015geometric}.
It does limit the possibility of predicting the properties of physical materials though, and therefore is a significant obstacle to realizing the vision of integrated computational materials design.

Methods that simulate the evolution of three-dimensional grain structures explicitly \cite{1990PhaseTransNagai,2000SIAMJourSciCompKuprat,2010ModelSimuMaterSciEngSyha,2011ActaMateLazar} are referred to as discrete boundary methods below.
One of the main difficulties faced by such methods is with maintaining a consistent mesh of the grain boundary surfaces during a topological transition.
Indeed, there did not even appear to be a way to enumerate a broad class of possible topological transitions in a general grain boundary network until quite recently \cite{2021PRMEren}, and it is not at all obvious how to explicitly implement such transitions without at least knowing what they are.
Nevertheless, discrete boundary methods do offer several distinct advantages with respect to microstructure modeling.
They often require far fewer mesh points than their diffuse counterparts, offering a dramatic reduction in runtime computational cost.
They also allow various defect properties, including grain boundary energies and triple line energies, to be explicitly defined in a way that is difficult with diffuse boundary methods.
Along these lines, Kuprat previously developed GRAIN3D to simulate grain growth \cite{2000SIAMJourSciCompKuprat} on a volumetric mesh with the gradient weighted moving finite element (GWFE) method, though the proposed topological transitions were not necessarily physical and could substantially affect the microstructure trajectory.
Shya and Weygand instead proposed a method to simulate grain growth on a surface mesh and handled topological transitions by decomposing them into sequences of elementary operations \cite{2010ModelSimuMaterSciEngSyha}, but did not offer any assurance that such decompositions would not change the microstructure trajectory.
Lazar et al.\ \cite{2011ActaMateLazar} proposed a discretized formulation of the MacPherson-Srolovitz relation \cite{2007NatureMacPhersonSrolovitz} to simulate ideal grain growth on a surface mesh.
While this only required that a small number of topological transitions be implemented, the explicit assumption of isotropic grain boundary properties precluded simulations of more general systems.

Two of the authors recently proposed a discrete interface method that addresses several of the computational challenges associated with explicit microstructure meshing, including a way to construct all possible topological transitions around a junction point and an energetic criterion to select one to apply \cite{2021PRMEren}.
The implementation \cite{2021VDlib} is based on SCOREC \cite{2016ACMIbanez} and uses a volumetric microstructure mesh, potentially allowing the addition of other necessary physics to build a general framework for realistic simulations of microstructure evolution.
SCOREC is an open source, massively parallelizable finite element framework with the adaptive meshing capabilities that are necessary to reach representative material volumes and to efficiently maintain the mesh element quality and desired degree of refinement.
SCOREC is specifically able to improve the quality of low-quality elements by local remeshing operations that minimally disturb the embedded surface mesh and make the computational expense of many operations, e.g., collapsing an individual grain, constant with respect to the system size.
The remeshing operations can also be used to refine a microstructure mesh.
For example, a polycrystalline microstructure consisting of Voronoi polyhedra can be converted into a microstructure mesh by initially placing a single vertex on the interior of each boundary line, boundary surface, and grain volume, and subsequently refining using the mesh adaptation capabilities of SCOREC.
This work offers an initial comparison between this discrete interface method (detailed further in Ref.\ \cite{2021PRMEren}) and a more well-established phase field approach.

This paper introduces a set of three test cases to evaluate the relative accuracy and numerical cost of simulations of grain boundary motion, and uses this set to compare the discrete interface and phase field methods.
The three cases correspond to several of the simplest configurations involving the motion of a grain boundary surface, a triple junction (TJ), and a quadruple point (QP).
The grain boundary properties are assumed to be isotropic; a coarsening grain structure with isotropic grain boundary properties is said to be the ideal grain growth system, and provides a basis for the comparison of all other evolving grain structures.
Analytical forms for the evolving geometries are known for the spherical surface and TJ cases \cite{mullins1956,1999InterfaceSciGottstein}, and the TJ and QP configurations have well-defined steady-state geometries.
It is also of interest whether the two methods converge to the same geometries in situations for which analytical solutions are not known, since there is likely no other way to verify the simulations in such cases.
While several of these configurations have been studied before, they are not usually considered in conjunction despite the benefits of doing so.
Namely, the increasing complexity of the grain boundary configurations among the three test cases introduces different sources of systematic error to the grain boundary motion, and these errors can be more easily identified by comparing the test cases to one another.

It is desirable to establish the nature of any systematic errors and the accuracy of the simulation methods for a system with isotropic grain boundary properties before attempting to do so with more general grain boundary energy and mobility functions.
The two methods considered in the present work will be capable of simulating the motion of grain boundaries with anisotropic properties when such functions become available.
The discrete boundary method uses equations of motion that allow for general grain boundary properties and grain boundary lines that join an arbitrary number of grain boundaries \cite{2017ActMateMason}.
The multiphase field model was developed to simulate the faceting of grain boundaries with energies that depend on boundary plane orientation, though this requires calculating a fourth-order derivative of the order parameters \cite{2019ModSimMatSciRibot}.

This paper is structured as follows.
We begin with a discussion of the discrete interface method and its implementation, followed by an analogous discussion for the phase field/diffuse interface counterpart.
We then apply both methods to a set of three test cases: a two grain system (shrinking sphere), a three grain system (triple junction), and a five grain system (quadruple point).
The behavior of the discrete and diffuse models is compared for each of the examples {\it vis-\`a-vis} analytic predictions and the models' internal length scales.
The performance of the discrete interface model is briefly discussed, and then we conclude with a general discussion of the behaviors of the two models and a summary of recommendations for best practice in discrete interface modeling.

\section{Methods}
\label{sec:methods}

Assuming that grain boundary properties are independent of grain boundary crystallography implies that the grain boundary network evolves along the negative gradient of the total boundary area. This is usually expressed by means of the Turnbull equation \cite{1951PROGINMETALPHYSICSBurkeTurnbull}
\begin{equation} \label{eqn:Turnbull}
  \bm{v} = m \gamma K \nvec{n}
\end{equation}
governing the motion of each boundary patch where $\bm{v}$ is the velocity, $m$ and $\gamma$ are the mobility and energy per unit area, $K$ is the mean curvature (the sum of the principle curvatures), and $\nvec{n}$ is the unit normal vector.

While this is sufficient to determine the time evolution of a closed surface, the Turnbull equation does not specify what happens at the TJs or QPs of the grain boundary network.
One of the essential differences between discrete and diffuse interface models is the governing equations for precisely these locations. 
Discrete interface models generally represent the TJs and QPs as distinct entities with explicit geometries, and sometimes provide additional governing equations specific to these locations \cite{2011ActaMateLazar}.
This is in contrast to the implicit approach of most diffuse interface methods which do not track TJs or QPs explicitly (while some diffuse interface methods do include higher order terms to account for the distinct behavior of line or point defects, these can come at extreme computational cost).
Each surface instead evolves according to the Turnbull equation with the geometric singularities at the TJs and QPs regularized by the diffuse interfaces.
This difference in the handling of TJs is significant since the TJs define the geometric conditions at a grain boundary's edges, thereby constraining the evolution of the grain boundary surface and likely the overall microstructure trajectory.
It is for this reason that the angles between adjoining grain boundary surfaces are often used as simple scalar measures of the simulation accuracy in \cref{sec:results} below.

\subsection{Discrete interface model}
\label{subsec:discrete_interface}

As implied by the name, every discrete interface model uses a discrete representation of the grain boundary network.
A discrete representation entails that the grain boundary network geometry is represented by a collection of simple geometric objects, or elements, along with a description of how to join those elements together.
The result is known as a surface mesh in three dimensions, and can be advantageously extended to a volumetric mesh to provide a discrete representation of the grain interiors as well.
VDlib \cite{2021PRMEren, 2021VDlib} is a C++ library based on SCOREC \cite{2016ACMIbanez} that represents a grain structure by means of a volumetric mesh containing tetrahedra, triangles, edges, and vertices.

\begin{figure}
\centering
\includegraphics[width=5.5cm]{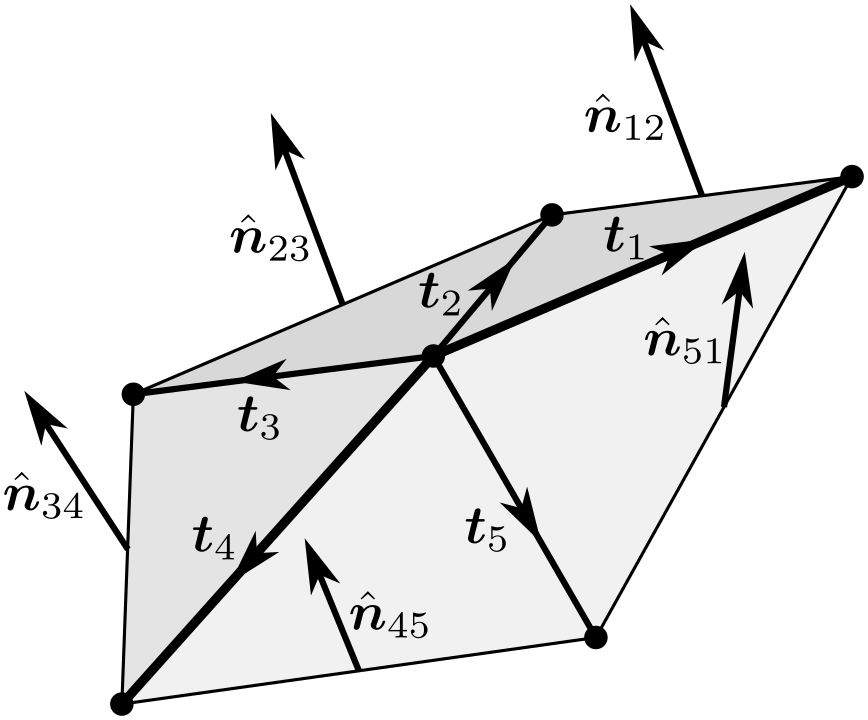}
\caption{Vectors describing the geometry around a vertex of the surface mesh. The central vertex is connected to five edges $\bm{t}_i$ and five triangles with unit normal vectors $\nvec{n}_{ij}$. The TJ along the edges $\bm{t}_1$ and $\bm{t}_4$ is shown in bold.}
\label{fig:dihedral_angle}
\end{figure}

There are two operations involved in updating the mesh to evolve the microstructure.
The first moves the vertices of the mesh according to established equations of motion \cite{2017ActMateMason} that allow for anisotropic surface energies and arbitrary drag coefficients
(the counterpart to the usual grain boundary mobility).
The idea is that the velocity $\bm{v}$ of any given vertex should be such that the driving force $\bm{F}$ on the vertex is precisely balanced by the sum of drag forces $\bm{D} \bm{v}$ resulting from the motion of the adjoining grain boundary elements, where $\bm{D}$ is the drag tensor and $\bm{v}$ is the grain boundary velocity.
The capillary force acting on the vertex is given by
\begin{align}
  \bm{F} &= \sum_i \nvec{t}_i \tau_l(\nvec{t}_i) +  
  \frac{1}{2} ||\bm{t}_i|| \sum_{j:\{i,j\}\in \Delta} (\nvec{n}_{ij} \times \nvec{t}_i) \gamma (\nvec{n}_{ij}) + \nvec{n}_{ij} \left.\frac{\partial \gamma}{\partial \phi_i}\right|_{\nvec{n}_{ij}},
\end{align}
where $\tau_l$ and $\gamma$ are the line and surface energy functions, $\bm{t}_i$ is the vector along edge $i$ starting at the vertex and $\nvec{t}_i$ is the corresponding unit vector, $\nvec{n}_{ij}$ is the normal of the triangle formed by edges $i$ and $j$, $j:\{i,j\} \in \Delta$ indicates an edge $j$ starting at the vertex such that edges $i$ and $j$ span a triangle $\Delta$, and $\phi$ defines the surface orientation around edge $i$;
\cref{fig:dihedral_angle} shows several of these quantities for a generic vertex of a surface mesh.
At force equilibrium the capillary forces are balanced by the drag forces $\bm{D} \bm{v}$ of the moving boundaries with
\begin{align}
  \bm{D} &=
    \delta_0 \bm{I} +
    \frac{1}{2} \sum_{i} \delta_1(\nvec{t}_i) 
    ||\bm{t}_i|| (\bm{I} - \nvec{t}_i \otimes \nvec{t}_i) +
    \frac{1}{6} \sum_{i, j \in \Delta} \delta_2(\nvec{n}_{ij}) 
    ||\bm{t}_i \times \bm{t}_j || (\nvec{n}_{ij} \otimes \nvec{n}_{ij})
\end{align}
where $\delta_{k}$ is the drag term associated with the $k$-dimensional simplicial boundary element.
The resulting boundary vertex velocity $\bm{v}$ is given by
\begin{equation}
  \bm{v} = \bm{D}^{-1} \bm{F}.
\end{equation}
One advantage of this formulation is that the motion of every boundary vertex is governed by the same equation, including those on the interiors of surfaces, along TJs, and at QPs.
If the point and line drag terms are zero, $\bm{D}\bm{v}$ reduces to the sum of the drag forces exerted by the neighboring triangles along the triangle normal directions for a given velocity $\bm{v}$.
Moreover, if the grain boundary properties are constant, then $\delta_2 = 3 / m$ and this further reduces to a discrete version of \cref{eqn:Turnbull} with an accuracy that depends on the product of the edge length and the mean curvature of the surface.
% The explicit nondimensionalized forms of the above equations are given in \ref{app:nondim}.

% If the point and line drag terms are zero, $\bm{D}\bm{v}$ reduces to the tensor operator that gives the sum of the drag forces exerted by the neighboring triangles the along triangle normal directions for a given velocity $ \bm{v} $.
% For a surface vertex, taking $\delta_2$ as the inverse of mobility $m$, the inverse of the drag matrix multiplied with the unit vector along the force twice, multiplied by one third of the area of the neighboring triangles is roughly equal to the mobility.

Apart from the motion of the mesh vertices, the accuracy of the discrete interface model is highly dependent on the element quality, where low-quality elements do not resemble equilateral triangles or tetrahedra \cite{1994parthasarathy,2000field}.
Without regular intervention and adaptation of the mesh, the quality of mesh elements generically degrades with grain boundary motion, even to the point of elements inverting.
The discrete interface method handles this by using MeshAdapt \cite{2005CompMetAppMechEngLi} to locally remesh where the element quality falls below a threshold value, and coarsening or refining edges with lengths below or above threshold values.
The target edge length $\ell_e$ is constant in time and space for any given simulation, and an edge is coarsened or refined if the edge length $l$ is outside the interval $0.7 \ell_e \leq l \leq 1.5 \ell_e$.
These operations are used sparingly though, since apart from the computational expense local remeshing can perturb the grain boundary geometry.
Specifically, these operations are the source of the discontinuous jumps observed in the discrete interface model results in \cref{sec:results}.
% In addition, the quality improvement operation is currently applied by default after the refinement operation and when these operations compete the mesh may get perturbed twice each time the adaptation is applied.
% These lead to seesaw-like behavior that are observed as in Fig. \ref{subfig:discrete_ang_trip_1} which are reduced as the refinement increases.

\subsection{Diffuse interface model}
\label{subsec:diffuse_interface}

Comparison to a standardized diffuse interface model provides verification of the discrete interface model.
In this work we apply the multiphase field model implemented following the presentation in Refs.\ \cite{moelans2008quantitative_1,moelans2008quantitative_2} which are general references for this section.
A brief overview is provided here.
For a system in a region $\Omega\subset\mathbb{R}^3$ with $N$ grains, $N$ order parameters (denoted as the vector of functions $\bm{\eta}=\{\eta_1,\ldots,\eta_N\}\subset C_2(\Omega)$) are defined such that the region occupied by the $i$th grain is precisely the support of $\eta_i$.
The free energy of the system is then defined to be
\begin{align}\label{eq:pf_free_energy}
  W[\bm{\eta}] = \int_\Omega \Big(w(\bm{\eta}) + \frac{1}{2}\sum_{n}k\,|\nabla\eta_n|^2\Big)\,d\bm{x},
\end{align}
where $w$ is the chemical potential and $k$ is a model parameter to be discussed subsequently
(the use of functional brackets should be understood to indicate dependence on the argument and any temporal or spatial derivatives).
The following polynomial form is used for the chemical potential:
\begin{align}
  w(\bm{\eta}) = \mu \sum_n\Big(\frac{1}{4}\eta_n^4 - \frac{1}{2}\eta_n^2 + \frac{3}{4}\sum_{m>n}\eta_m^2\eta_n^2\Big), \ \ \ \mu=3.26.
\end{align}
The coefficient for the boundary term is related to the grain boundary energy $\gamma$ by
\begin{align}
  k = \frac{3\ell_{GB}}{4}\gamma,
\end{align}
where $\ell_{GB}$ is the diffuse boundary width.
The evolution of $\bm{\eta}$, which determines the overall evolution of the microstructure, follows an $L^2$ gradient descent to minimize \cref{eq:pf_free_energy}.
The resulting kinetic evolution equation, expressed in terms of the variational derivative, is
\begin{align}
\label{eq:variational}
  \frac{\partial\eta_n}{\partial t} = -L\frac{\delta W}{\delta\eta_n},
\end{align}
where the rate coefficient $L$ is related to the traditional boundary mobility $m$ by
\begin{align}
  L = \frac{4}{3}\frac{m}{\ell_{GB}}.
\end{align}

Phase field simulations are often computationally costly, and this has led to a variety of methods to accelerate them.
Spectral methods can result in a substantial performance increase \cite{chen1998applications,chen2002phase,tourret2022phase}, though this comes at the cost of limited resolution of fine features and the restrictive requirement that the computational domain be periodic.
Real-space (non-spectral) methods instead require strategic meshing techniques, such as adaptive mesh refinement (AMR) \cite{tourret2022phase}, to avoid prohibitively excessive mesh size.
However, as with the discrete interface method, they can be easily implemented in non-periodic systems and systems with complex geometry.
Therefore, real-space methods with adaptive mesh refinement are the most appropriate benchmark against which to compare the present discrete interface method.

In this work, all diffuse boundary calculations are performed using Alamo, a high performance multiphysics code that uses block-structured adaptive mesh refinement (BSAMR) with a strong-form elasticity solver to perform diffuse interface calculations \cite{runnels2021massively}.
Alamo is built on the AMReX package, developed by Lawrence Berkeley National Laboratory \cite{zhang2019amrex}.
All of the results presented here were run on a desktop computer and generally completed in less than an hour depending on the chosen parameters.
Of particular interest is the convergence of the solution with respect to the boundary width, $\ell_{GB}$, which determines the diffuse boundary length scale.
The exact solution is recovered as $\ell_{GB}\to 0$, but this comes at the expense of increased computational cost.
In this work we are particularly interested in the relationship between $\ell_{GB}$ and the discrete interface model counterpart.

\subsection{Topological transitions}

\begin{figure}
\center
\subfloat[]{%
	\label{subfig:topo_init}{%
		\includegraphics[width=0.2\linewidth]{%
			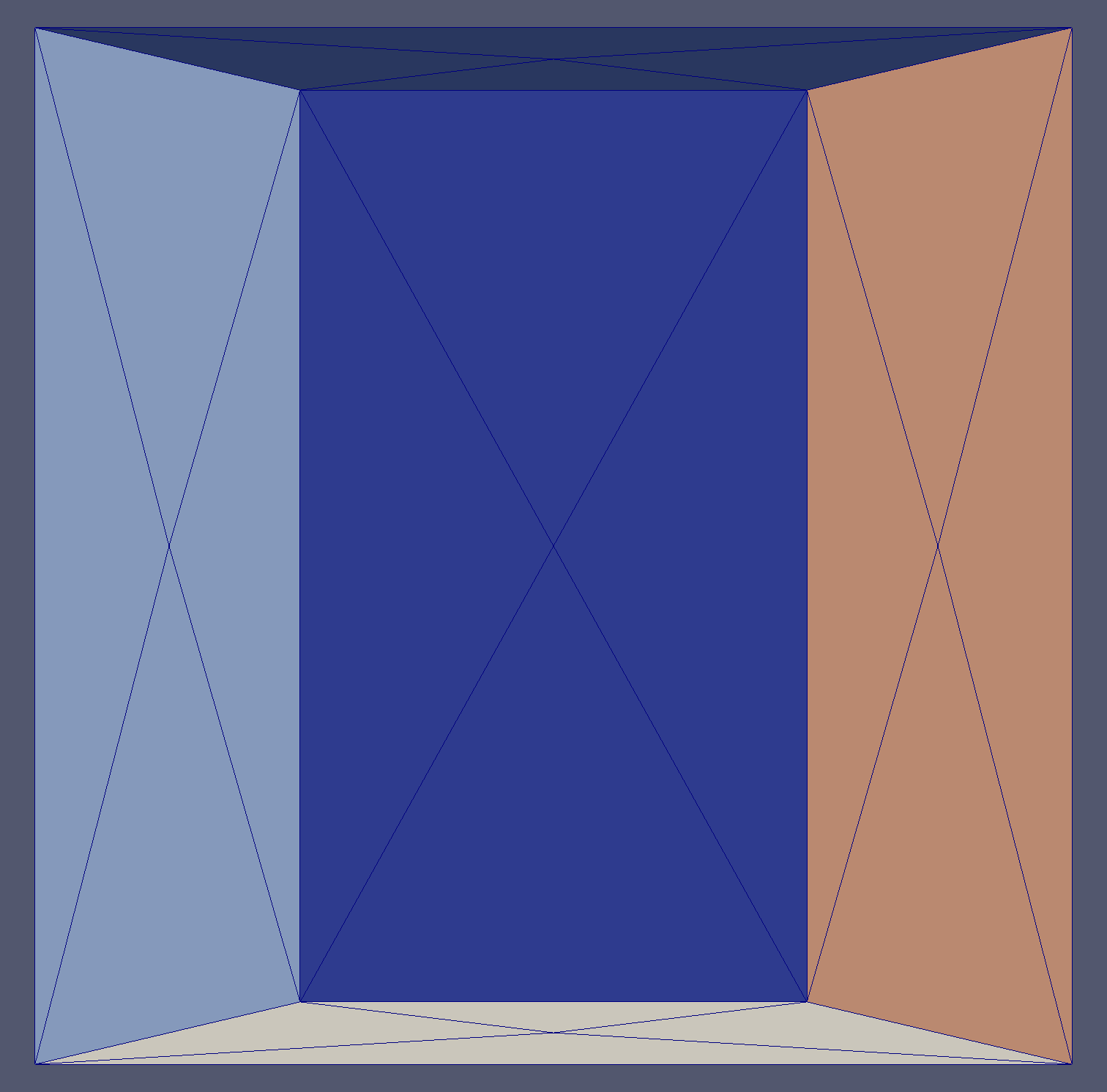}}}
\subfloat[]{%
	\label{subfig:disc_trans1}{%
		\includegraphics[width=0.2\linewidth]{%
			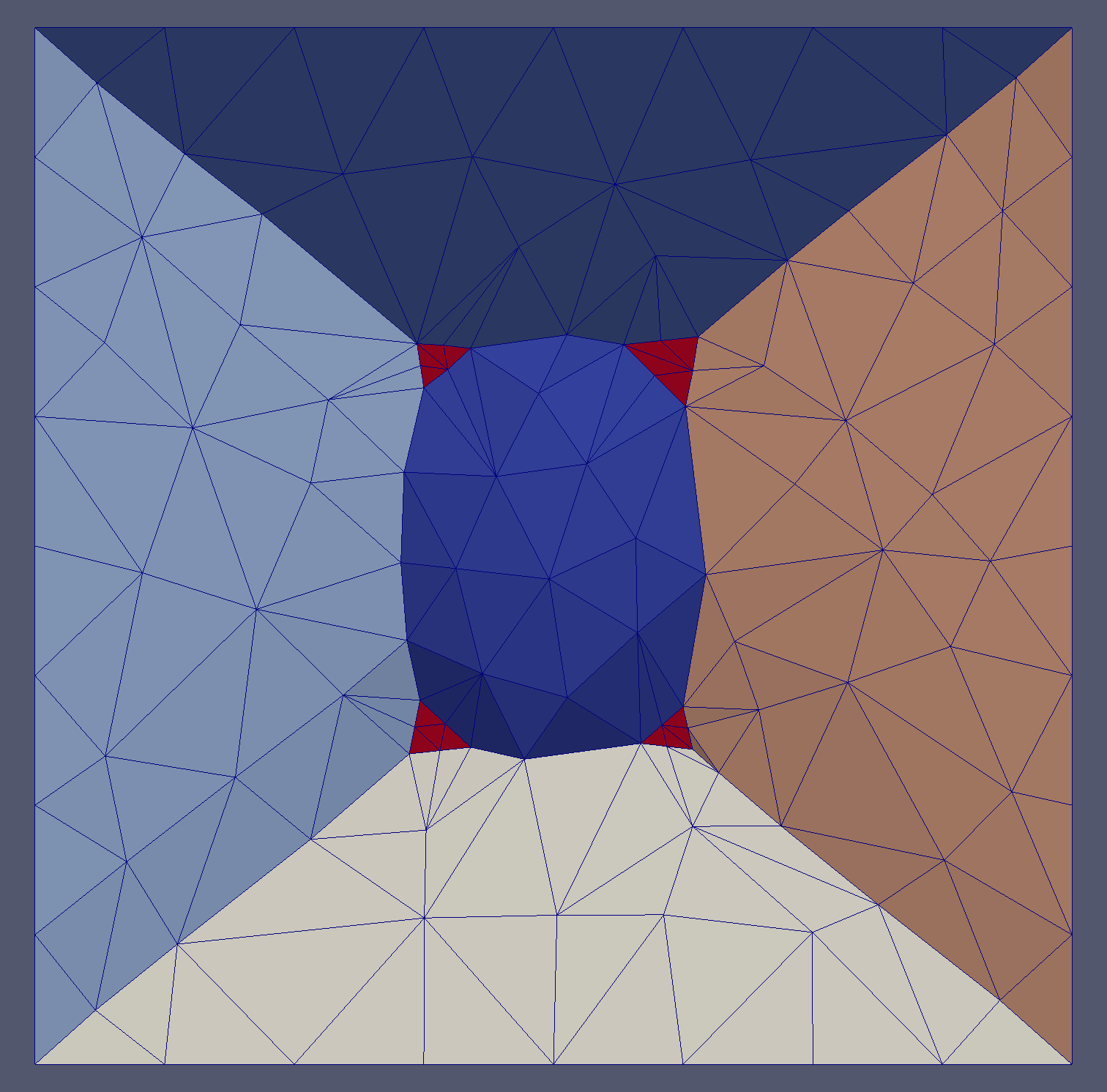}}}
\subfloat[]{%
	\label{subfig:disc_trans2}{%
		\includegraphics[width=0.2\linewidth]{%
			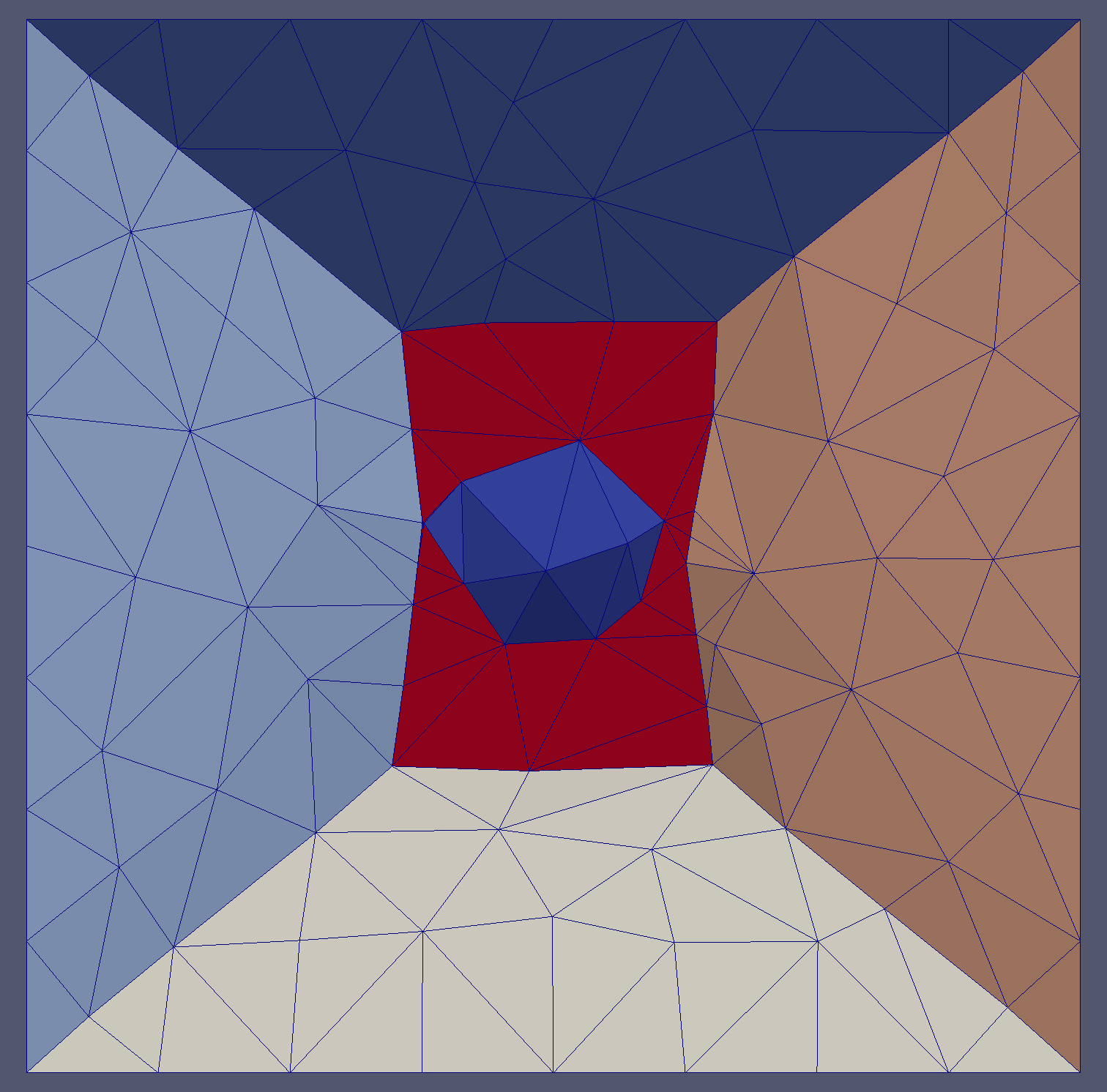}}}
\subfloat[]{%
	\label{subfig:disc_trans3}{%
		\includegraphics[width=0.2\linewidth]{%
			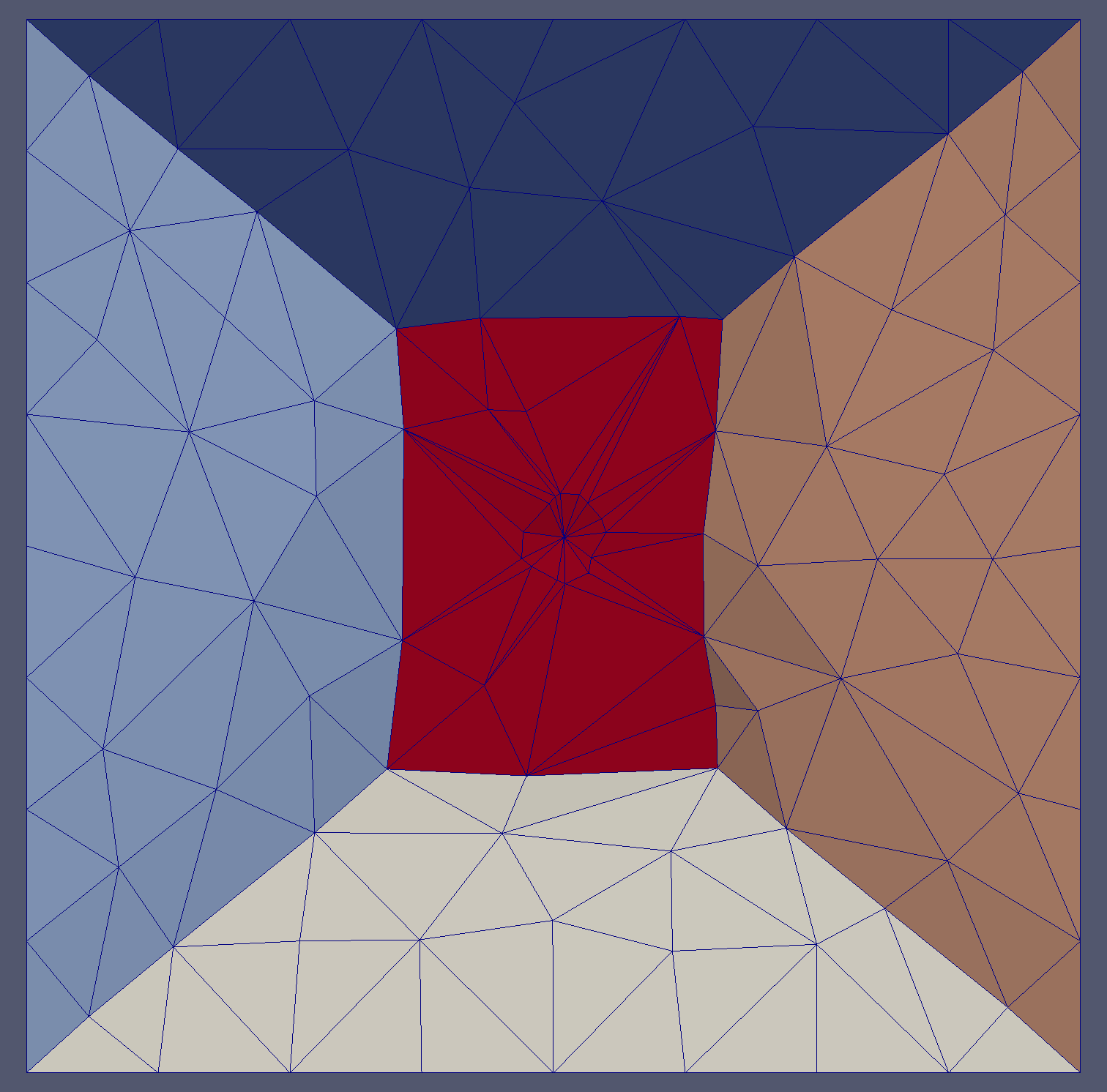}}}
\subfloat[]{%
	\label{subfig:disc_trans4}{%
		\includegraphics[width=0.2\linewidth]{%
			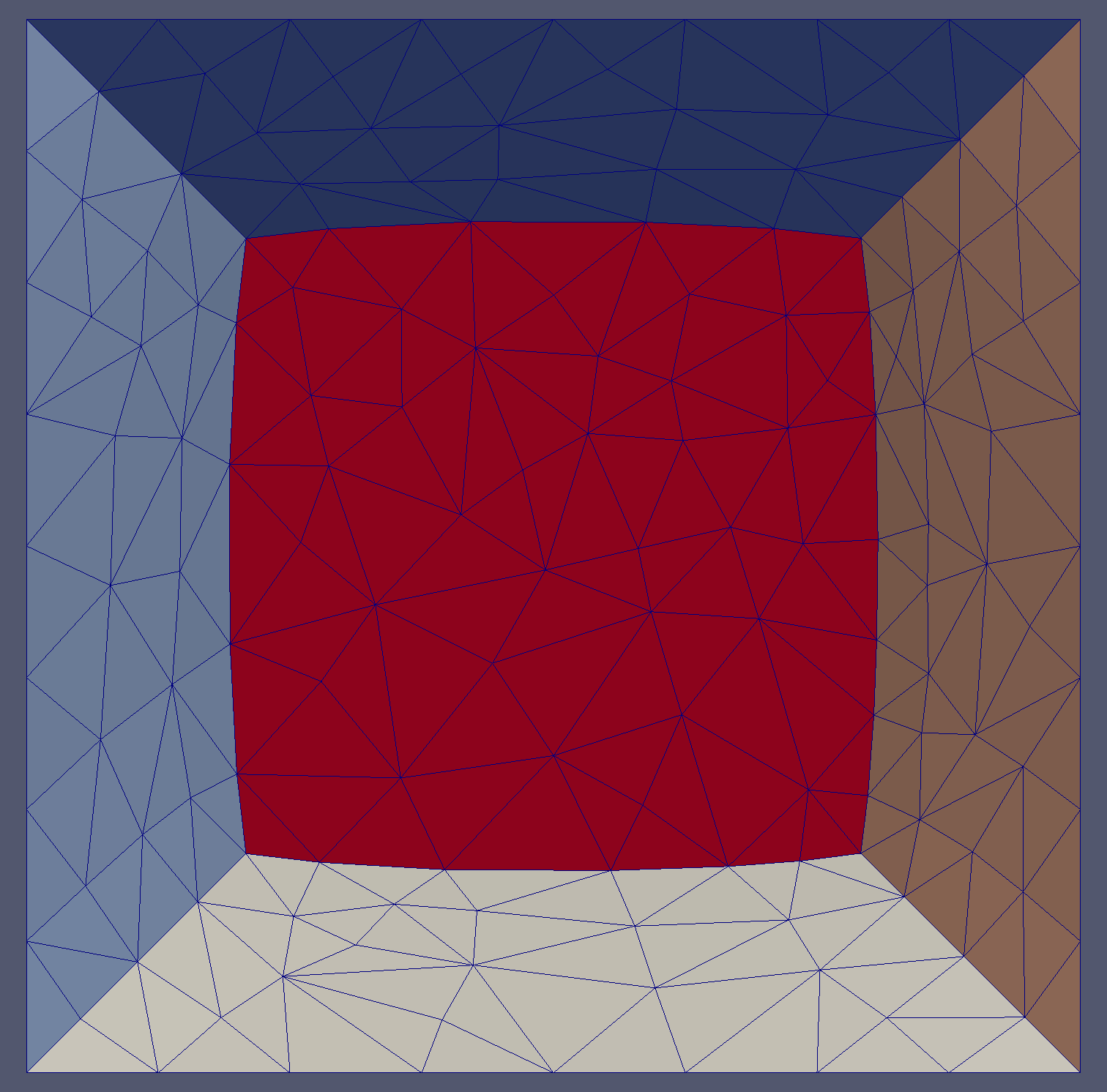}}}

\subfloat[]{%
	\label{subfig:topo_init_pf}{%
		\includegraphics[width=0.2\linewidth]{%
			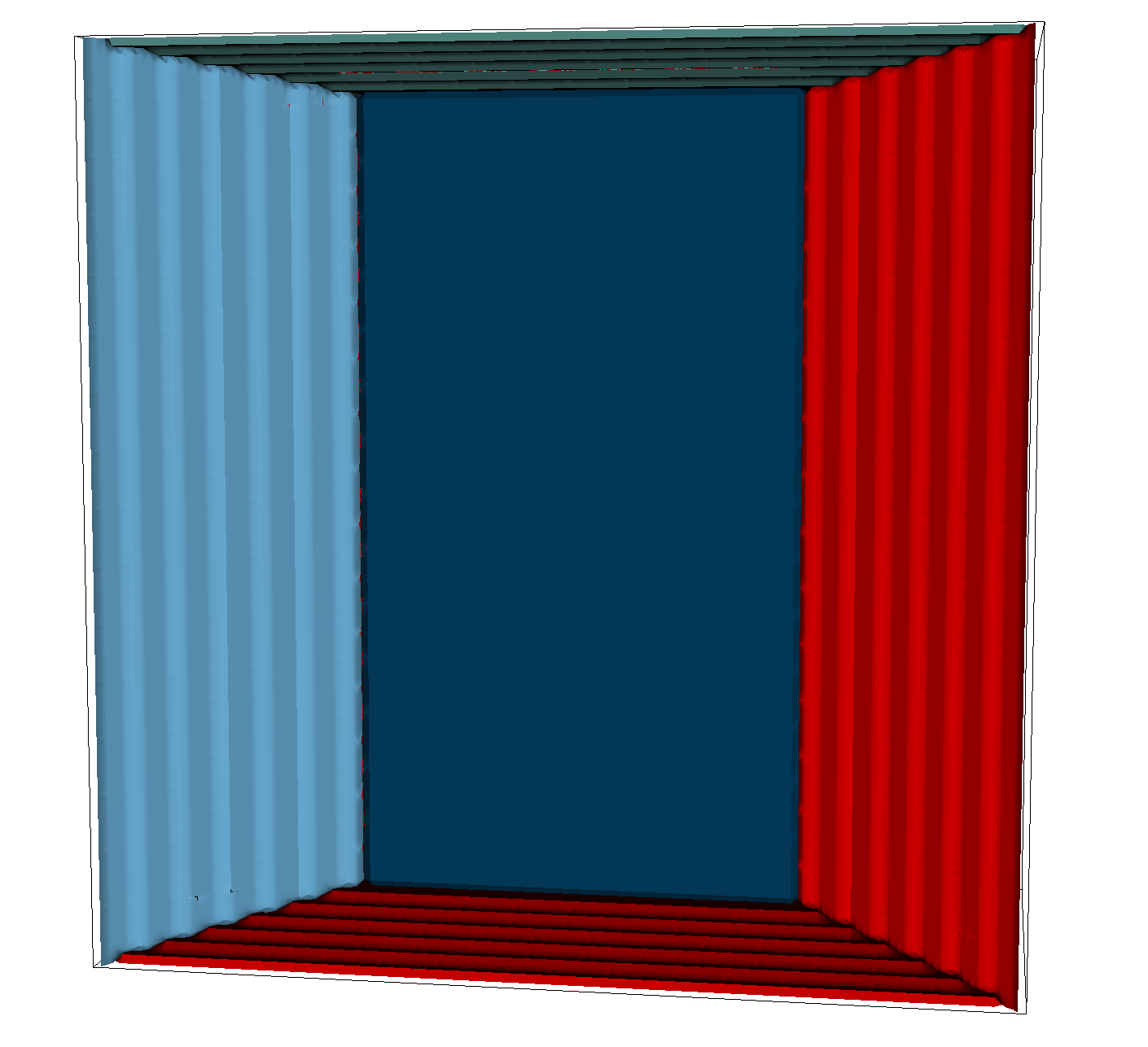}}}
\subfloat[]{%
	\label{subfig:disc_trans1_pf}{%
		\includegraphics[width=0.2\linewidth]{%
			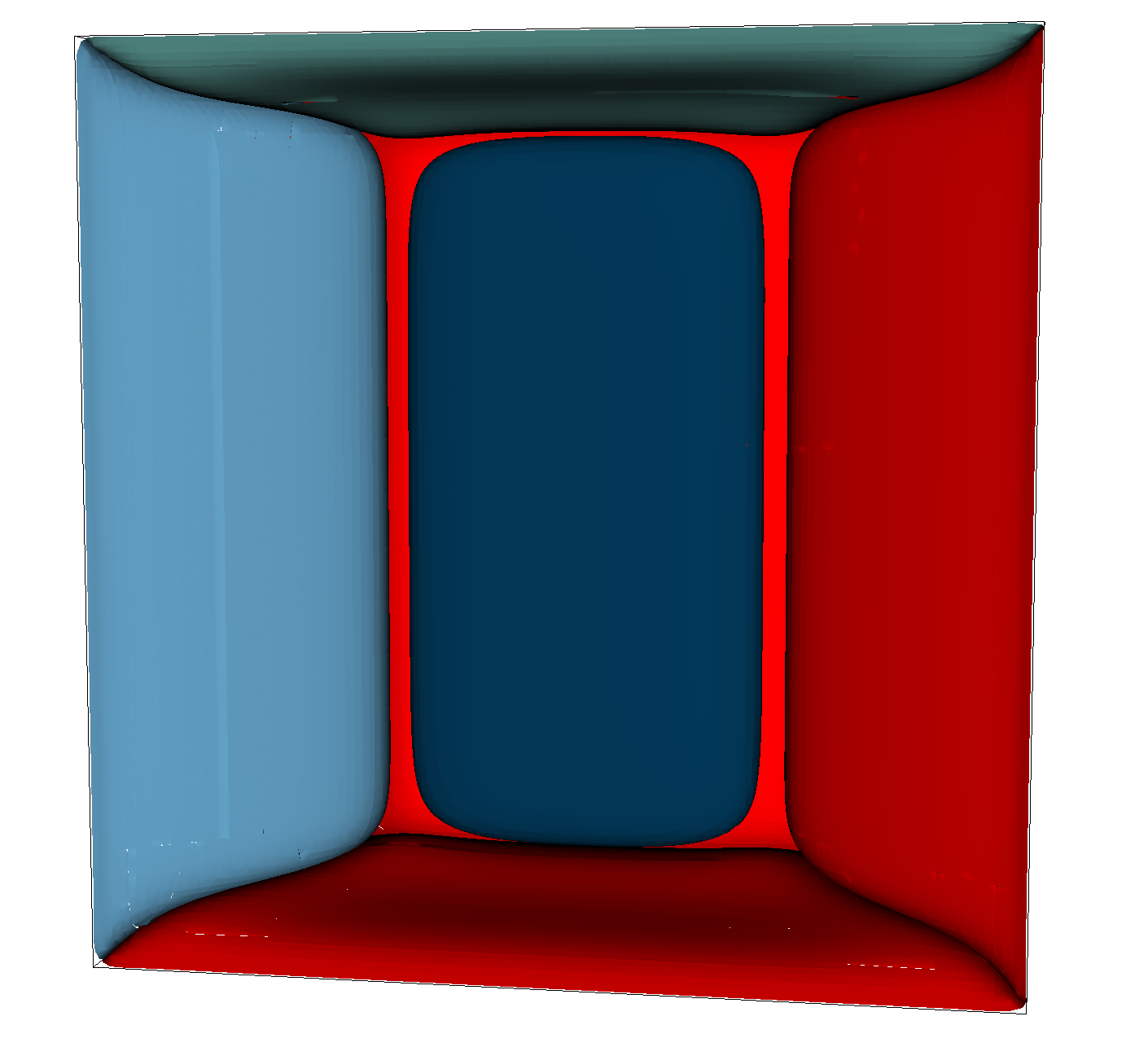}}}
\subfloat[]{%
	\label{subfig:disc_trans2_pf}{%
		\includegraphics[width=0.2\linewidth]{%
			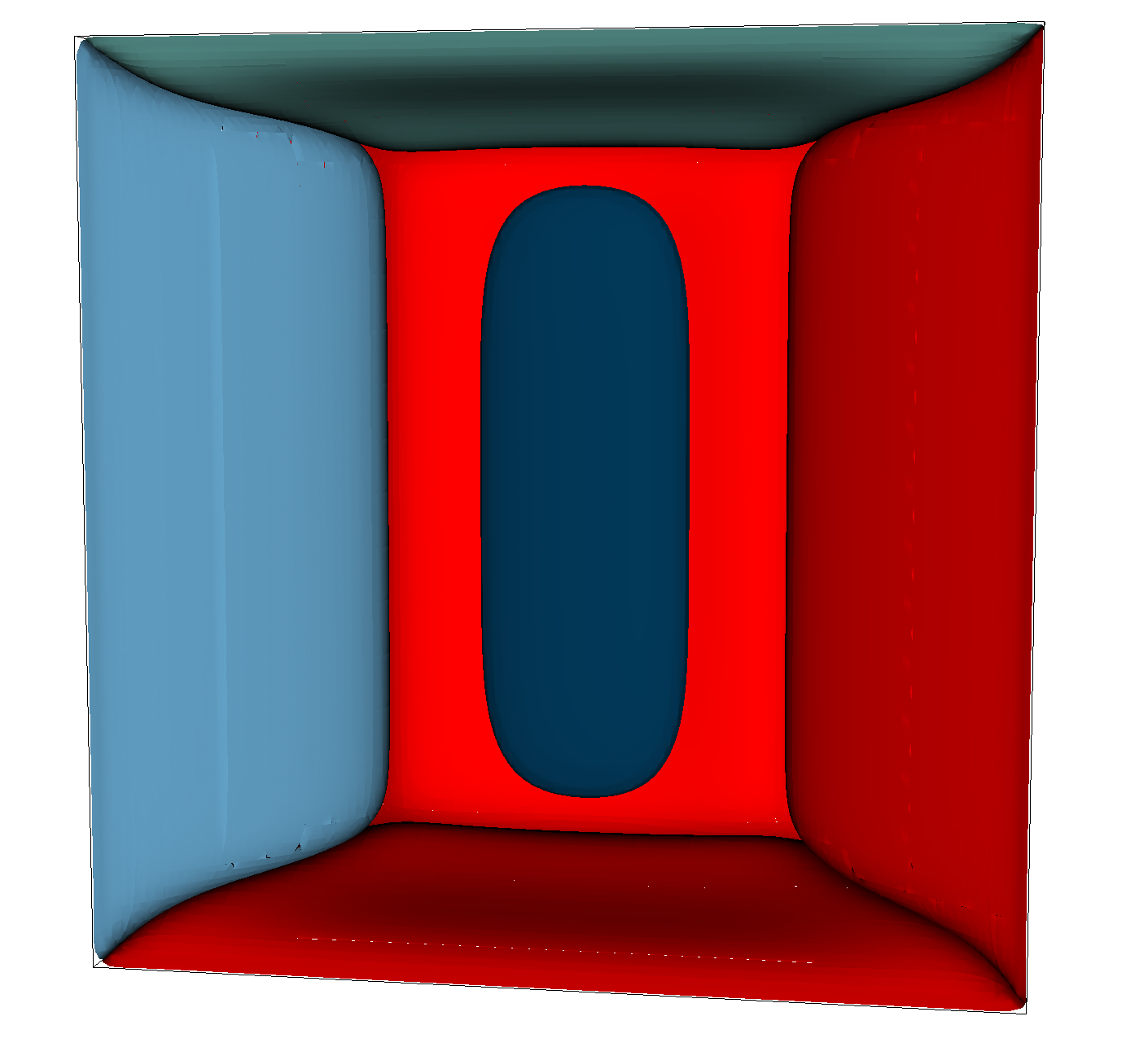}}}
\subfloat[]{%
	\label{subfig:disc_trans3_pf}{%
		\includegraphics[width=0.2\linewidth]{%
			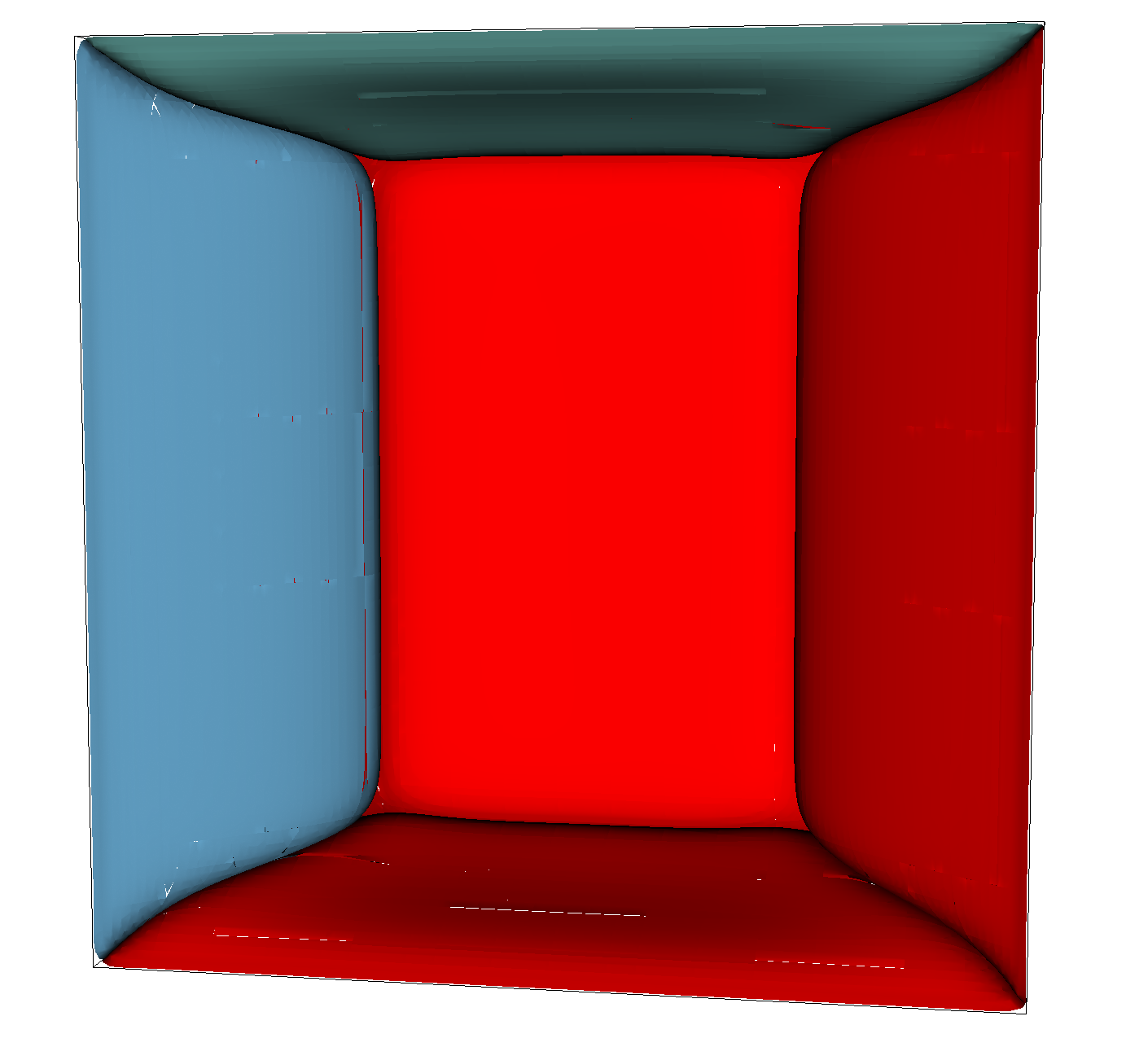}}}
\subfloat[]{%
	\label{subfig:disc_trans4_pf}{%
		\includegraphics[width=0.2\linewidth]{%
			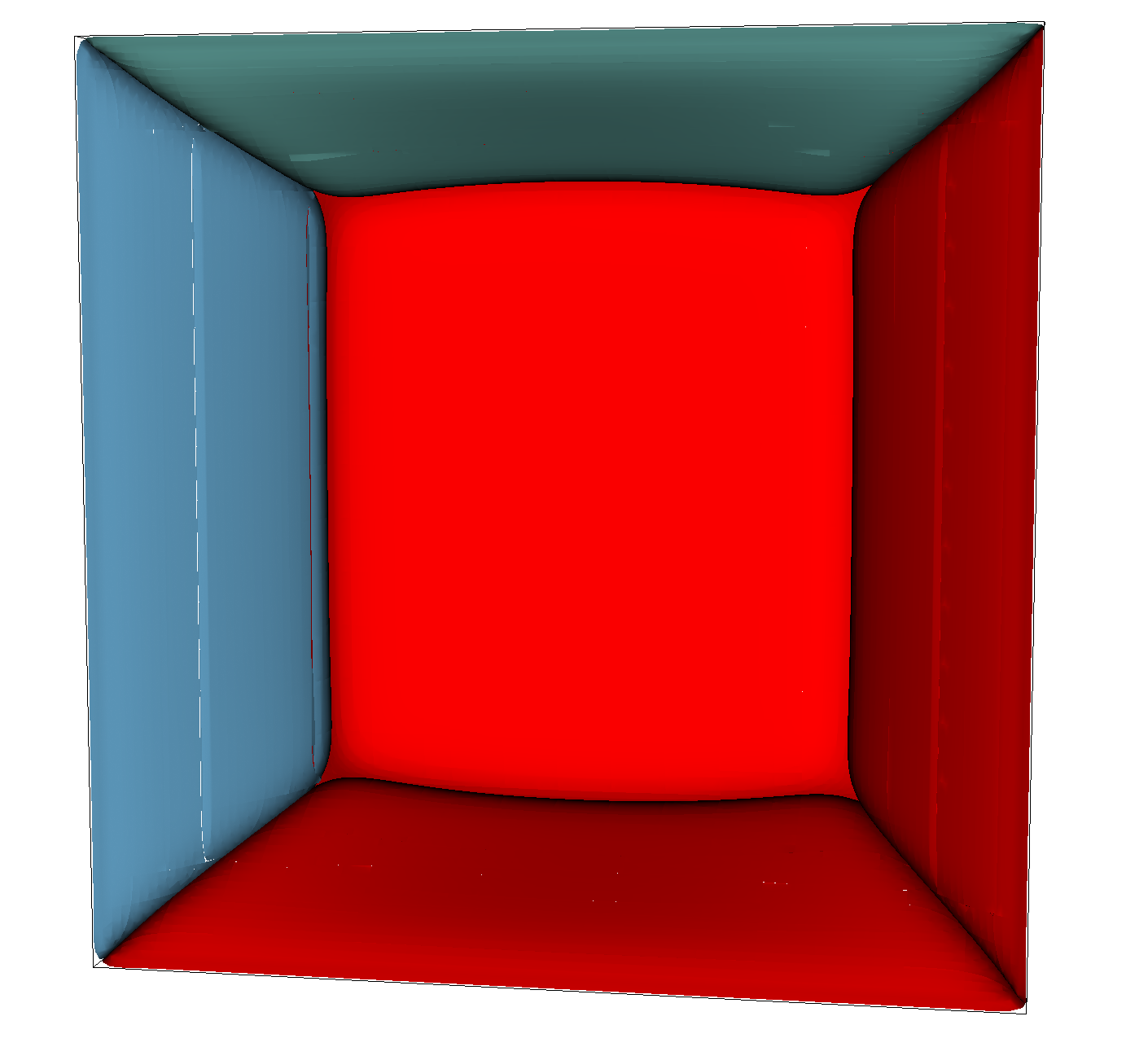}}}

\caption{\label{fig:topology}Geometric and topological changes in a structure where the central grain is initially a rectangular prism surrounded by six grains.
(a)-(e) and (f)-(j) show the structure after corresponding elapsed times in the discrete and diffuse interface models, respectively.}
\end{figure}

As stated in the introduction, one motivation for using diffuse interface methods for microstructure evolution is that the implicit nature of the grain boundaries allows topological transitions to occur without requiring that all possible transitions be explicitly enumerated.
The purpose of this section is to show that the challenge of enumerating and implementing such topological transitions for a discrete interface method is in fact surmountable \cite{2021PRMEren}.
This is accomplished by simulating the evolution of a non-generic grain structure that, despite the grain boundary properties being uniform and isotropic, involves topological transitions that are not generally handled by discrete interface methods \cite{2011ActaMateLazar,2000SIAMJourSciCompKuprat}.
The initial grain structure in Fig.\ \ref{fig:topology} contains a central rectangular prismatic grain surrounded by six other grains, the top one being removed for visual clarity.

For the discrete interface model on the top row, the initial topological transitions involve four triple lines collapsing into four triangular faces in Fig.\ \ref{subfig:disc_trans1};
this is a standard topological transition implemented in nearly all discrete interface methods.
The high symmetry of the initial condition subsequently results in the central grain detaching from the four side grains in Fig.\ \ref{subfig:disc_trans2}, with the four triangular faces that were previously introduced merging into an annulus around the central grain.
Such transitions and the resulting configurations would be difficult for other existing discrete interface methods.
While the method proposed by Syha and Weygand \cite{2010ModelSimuMaterSciEngSyha} could in principle handle such transitions, their assumption that junction lines are always bounded by junction points would be invalidated after the transition in Fig.\ \ref{subfig:disc_trans2}.
The central grain shrinks to the point of vanishing in Fig.\ \ref{subfig:disc_trans3}, and the structure has reached an effectively stable configuration in Fig.\ \ref{subfig:disc_trans4}.
Note that due to the anisotropy of the mesh and the adaptive remeshing perturbing the mesh slightly, the symmetrical transitions (e.g.\ collapse of the vertical triple lines just before Fig.\ \ref{subfig:disc_trans1}) did not occur exactly simultaneously.

The evolution of the same configuration in the diffuse interface model is quite different.
Each grain's boundaries were constructed as the surfaces where the value of the corresponding order parameter reached $0.5$
(the interaction of the underlying grid with the initial conditions produced the ridges visible in Fig.\ \ref{subfig:topo_init_pf}).
The central grain shrinks preferentially in the out-of-plane direction in Fig.\ \ref{subfig:disc_trans1_pf}, with four triangular faces appearing at the corners of the central grain shortly before the central grain completely separates from the adjacent grains in the horizontal direction.
That the geometric and topological evolution of the central grain should be different than in the discrete interface method is expected given the finite width of the diffuse boundaries.
Specifically, whenever two approaching boundaries are separated by a distance on the order of the boundary width, the gradients in the order parameter representing the two boundary interact, changing the effective boundary energy and mobility.
This effect is more than a postprocessing artifact of the surface reconstruction, and can change the microstructure trajectory in ways that resemble the differences in behavior between wet and dry foams \cite{drenckhan2015structure}.

While the discrete and diffuse interface methods converge to effectively the same configurations in this case (Figs.\ \ref{subfig:disc_trans4} and \ref{subfig:disc_trans4_pf}), microstructure trajectories are often unstable with respect to such perturbations in the proximity of a topological transition.
This phenomenon is beyond the scope of the current paper though, a detailed study of the evolution of surfaces, triple, points, and quadruple points in the absence of topological transitions (and as is performed here) being a necessary preliminary.

\section{Results and discussion}
\label{sec:results}

Three cases are considered in this section to quantify the systematic error of the discrete interface and phase field methods of simulating grain boundary motion.
The first is a spherical grain which evolves in a self-similar way; this is a standard configuration that is often used in the literature to verify that the Turnbull equation is obeyed in the absence of complicating factors \cite{1987Brakke,2000SIAMJourSciCompKuprat, 2012CompPhySaye,tonks2012object,2020CompMatSciFlorez}.
The second is a TJ that migrates along a semi-infinite grain boundary \cite{moelans2009comparative,2012CompPhySaye,2015CompMatSciJin,2018MatDsgnFausty}, eventually reaching a steady state configuration with a known profile and velocity \cite{mullins1956}.
The quantity considered below is the angle of the grain boundaries at the TJ, though in principle a stricter validation scheme could involve evaluating the simulation's ability to precisely reproduce the expected grain boundary geometry.
The third is a columnar hexagonal grain configuration that migrates along semi-infinite grain boundaries to allow a study of the steady state evolution of a QP \cite{2008ActaMateBarralesMora,2012CompPhySaye}.
Perhaps the reason this case appears less often in the literature is that an analytical solution for the boundary profile is not known;
instead, the angles between grain boundary traces on two cross-sections are evaluated for convergence and used to compare the two simulation methods.
The grain boundary geometries for the three test cases are described in their respective sections, have Neumann boundary conditions, and are constructed to make the grain boundary curvatures comparable.

It is expected that the accuracy of both the discrete and diffuse interface models will increase with decreasing internal length scale $\ell$, denoted as $\ell=\ell_e$ for the discrete model and $\ell=\ell_{GB}$ for the diffuse.
However, the accuracy cannot depend on any absolute length scale since then the accuracy could be improved simply by uniformly scaling the grain structure.
The accuracy therefore depends on $\ell$ relative to a second length scale that is characteristic of the evolving interface.
Since the accuracy should be invariant to the isometries of Euclidean space, the inverse of the interface's mean curvature is the natural candidate for the second length scale, and the accuracy of both models is expected to depend on the dimensionless product of $\ell$ and interface's mean curvature.
More precisely, all of the errors reported in this section are expected to be power laws in $\ell$, with the prefactor depending on the mean curvature and implementation details in a way that is difficult to parameterize (only the spherical grain has the same mean curvature everywhere).
For this reason, only the exponent of $\ell$ is generally reported in the following.

Many of the quantities reported below are nondimensionalized following the procedure in \ref{app:nondim} to facilitate the comparison of the discrete interface and phase field methods.
%\todo{EE: for analysis nondimensionalize the edge lengths, as well?}
A tilde indicates a nondimensionalized variable (with the exception of $\ell_e$ and $\ell_{GB}$ which are always nondimensionalized) and an analytical prediction is denoted by the subscript $t$, e.g., $\tilde{r}_t(\tilde{t})$ is the analytical prediction for the nondimensionalized radius of the sphere as a function of nondimensionalized time.
The equations of motion of the discrete interface method were integrated using a second order Runge--Kutta scheme with a maximum nondimensionalized time step of $1.2500 \times 10^{-5}$.

\subsection{Spherical grain}
\label{subsec:spherical_grain}

\begin{figure}
\center
\subfloat[]{%
	\label{subfig1:sph}{%
		\includegraphics[height=37mm]{%
			%./results_discrete/mesh_sph_crop.jpg}}}
			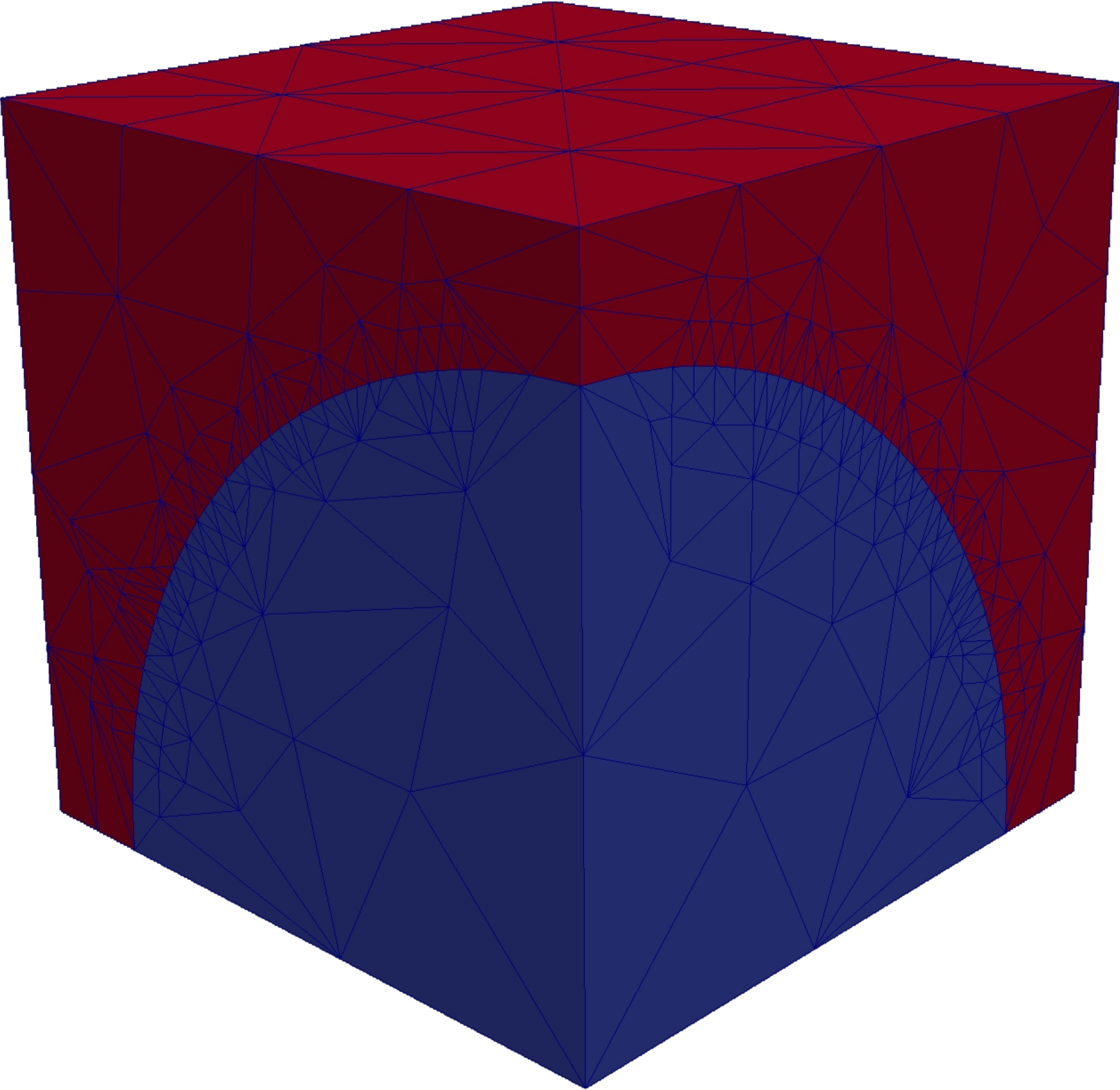}}}
%\hspace{-4em}
\subfloat[]{%
	\label{subfig1:sph_PF}{%
		\includegraphics[height=37mm]{%
			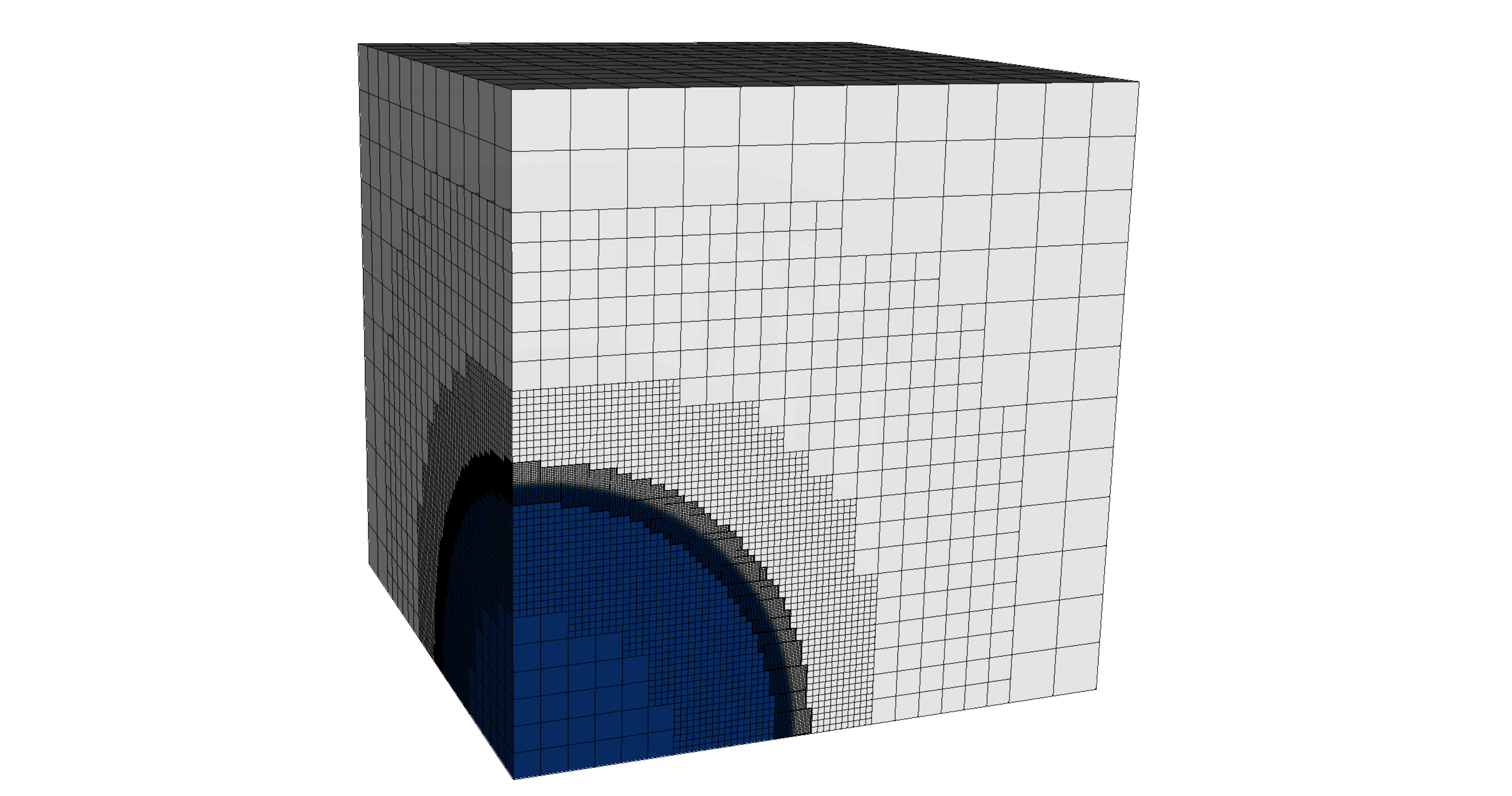}}}
\caption{\label{fig:sph_case}Initial geometries of the shrinking spherical grain within another grain for the (a) discrete and (b) diffuse interface methods.}
\end{figure}

The spherical grain case is intended to reveal the error when modeling surface motion in the absence of confounding effects from other grain boundary network components.
One advantage of this particular choice is that, provided the grain boundary properties are constant and isotropic, the evolution of a spherical grain is known analytically.
As derived in \ref{app:sphere}, the sphere shrinks uniformly with radius
\begin{equation} \label{eqn:roft}
  r_t(t) = \sqrt{r_0^2 - 4 m \gamma t}
\end{equation}
as a function of time.
Nondimensionalizing this equation reveals that a sphere starting with a radius of $\tilde{r}_t(\tilde{t}) = 1$ vanishes at $\tilde{t} = 0.25$.
The actual simulations deviate from \cref{eqn:roft} both because the initial geometries shown in \cref{fig:sph_case} are not precisely spheres and because the Turnbull equation in \cref{eqn:Turnbull} is not precisely followed, though these sources of error are reduced as the $\ell$ are made smaller.
Since the diffuse interface model doesn't perform well when the radius of the sphere approaches the grain boundary width, the magnitude of the error for the shrinking grain is quantified by the deviation of the sphere half-life $t_{half}$ from the analytical prediction $t_{half,t} = 3r_0^2/(16 m \gamma)$.
When nondimensionalized, this reduces to $\tilde{t}_{half,t} = 3 / 16$. 

\begin{figure}
  \includegraphics[width=\linewidth]{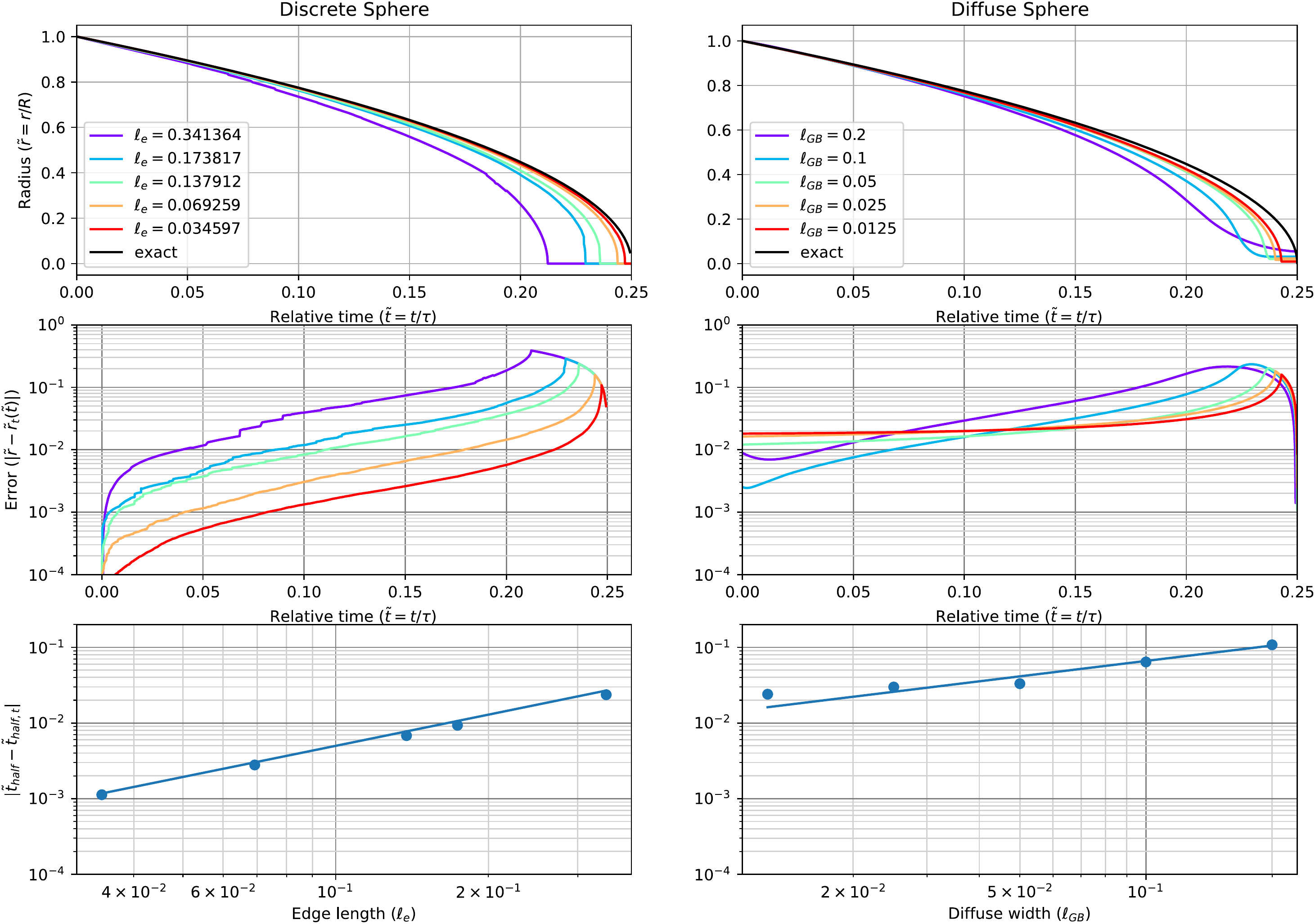}
  \caption{
    Comparison of shrinking spherical grain results for the discrete model (left) and the diffuse model (right); all quantities are nondimensionalized.
    (Top row) Plot of radius vs time, with color indicating the length scale and the exact solution in black.
    (Middle row) Plot of relative error in the radius vs time, with color indicating the length scale.
    (Bottom row) Plot of half-life error magnitude as a function of length scale.
  } \label{fig:sphere_r}
\end{figure}

Figure \ref{fig:sphere_r} shows the performance of the two models, with the discrete interface model on the left and the diffuse interface model on the right.
The top row shows the radius of the sphere as a function of time, where the color indicates the internal length scale and the exact solution is in black.
The roughness of the curves for the discrete interface model is due to remeshing to preserve the element quality, and the velocity in the diffuse interface model falls as the radius approaches the grain boundary width. 
The magnitude of the relative error in the radius as a function of time is shown in the middle row.
The error for the discrete interface model is caused by the magnitudes of the surface vertex velocities being larger than predicted by the analytical solution, perhaps as a consequence of the equations of motion being explicit and uncoupled.
That the accumulation of error accelerates with decreasing radius supports the hypothesis that the error generally depends on the product of $\ell_e$ and the mean curvature.
Meanwhile, there are likely two sources of error that contribute to the results for the diffuse interface model.
The error at early times is a postprocessing artifact that occurs when constructing isocontours to identify the location of the grain boundary, effectively resulting in an offset to the sphere radius.
The other source of error relates to the order parameter gradient at a grain boundary patch being affected by the presence of nearby patches.
This is most visible when the grain is about to collapse and grain boundary patches on opposite sides of the grain interact, reducing the gradient magnitude and the grain boundary velocity.
Conversely, the mean curvature of the surface causes neighboring grain boundary patches to interact, increasing the gradient magnitude and the grain boundary velocity at earlier times.
As with the discrete interface model, the magnitude of this effect at earlier times is proportional to the product of $\ell_{GB}$ and the mean curvature.

The bottom row of \cref{fig:sphere_r} shows the half-life error $|\tilde{t}_{half} - \tilde{t}_{half,t}|$ as a function of the internal length scale.
A conjugate gradient minimization algorithm and bootstrapping were used to fit $|\tilde{t}_{half} - \tilde{t}_{half,t}|$ to a power law in the internal length scale $\ell$.
This gives an exponent of $1.37 \pm 0.21$ for the discrete interface model and $0.678 \pm 0.085$ for the diffuse interface model, where the values are the medians and the uncertainties are half the interquartile range.
While the exponents could suggest that the error of the diffuse interface model decays slower than that of the discrete interface model with decreasing internal length scale, the errors in the apparent grain radius due to isocontour construction during postprocessing do not actually affect the microstructure trajectory.
This could motivate using the two-grain configuration with self-similar evolution analyzed by Mullins \cite{mullins1956} in the future since such postprocessing errors would likely not affect the long-time behavior.

\subsection{Triple junction}
\label{subsec:triple_junction}

\begin{figure}
\center
\subfloat[]{%
	\label{subfig1:trip}{%
		\includegraphics[height=37mm]{%
			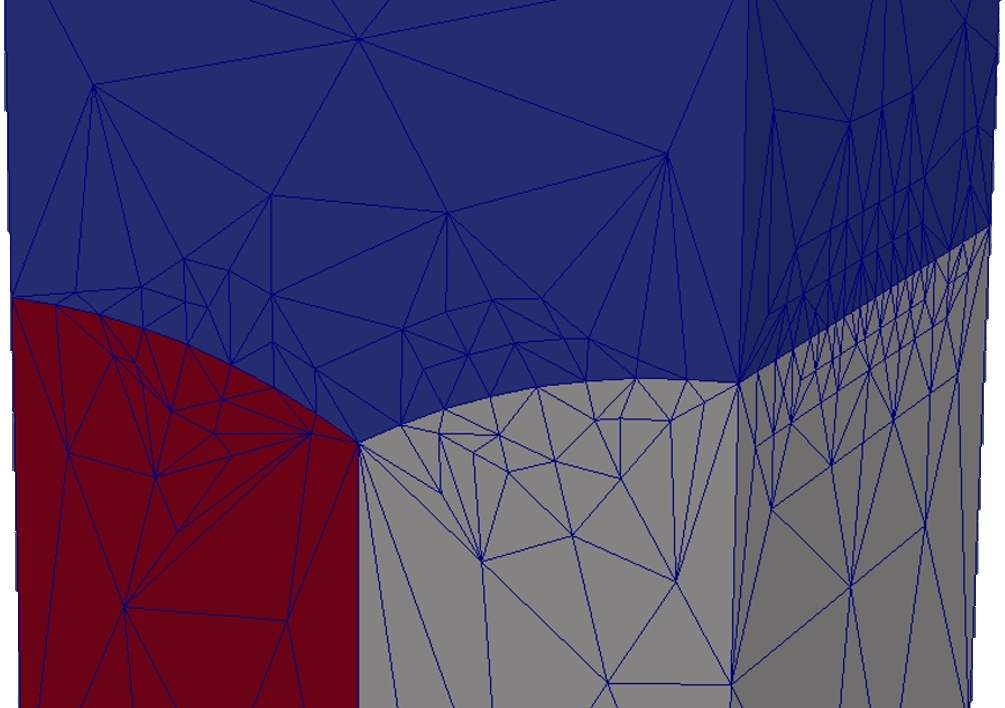}}}
\qquad
\subfloat[]{%
	\label{subfig1:trip_PF}{%
		\includegraphics[height=37mm]{%
			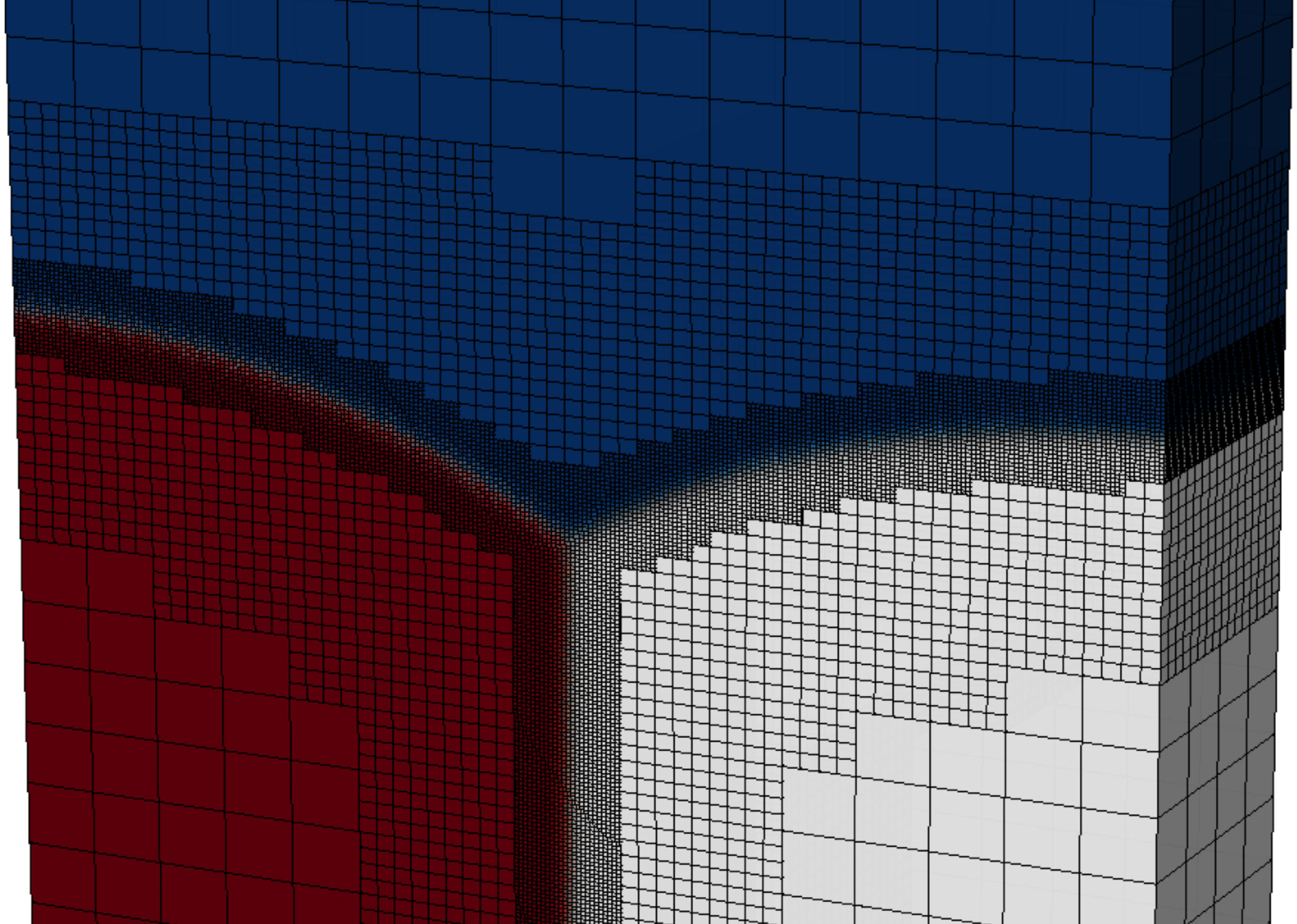}}}
\caption{\label{fig:trip_case}Initial geometries of the TJs for the (a) discrete and (b) diffuse interface methods. The structures have mirror boundary conditions in the lateral directions.}
\end{figure}

The purpose of the TJ case is to include a TJ in the moving boundary while keeping the grain configuration as simple as possible, ideally allowing the error of the equations of motion for the TJ to be identified by comparing the results to those for the spherical grain.
The initial geometries of the grain configuration are shown in \cref{fig:trip_case}, are constant in the out-of-plane direction, and have mirror boundary conditions in the lateral directions.
The rate of volume change of the top grain can be derived by applying the von Neumann-Mullins equation \cite{von1952,mullins1956} to the two-dimensional grain configuration in a plane perpendicular to the TJ.
Since there is one triple point per simulation cell in this plane, the rate of cross-sectional area change of the top grain per simulation cell width $L$ is $\pi m \gamma / 3$, and the rate of volume change of the top grain can be found by multiplying by the TJ length.
Mullins actually went further and solved for the steady-state profile of the moving boundary assuming constant and isotropic grain boundary properties \cite{mullins1956}.
If $x$ is distance from the left edge of the simulation cell and $y$ is height from the top of the red grain, then the steady-state profile of the grain boundary between the red and blue grains is
\begin{equation}
y(x) = -\ln[ \cos( \pi x ) ] / \pi.
\end{equation}
The width of the simulation cell as defined by the above equation would be $L = 1/3$, and is appropriately scaled to the actual dimensions of the simulation cell. 

\begin{figure}
  \includegraphics[width=\linewidth]{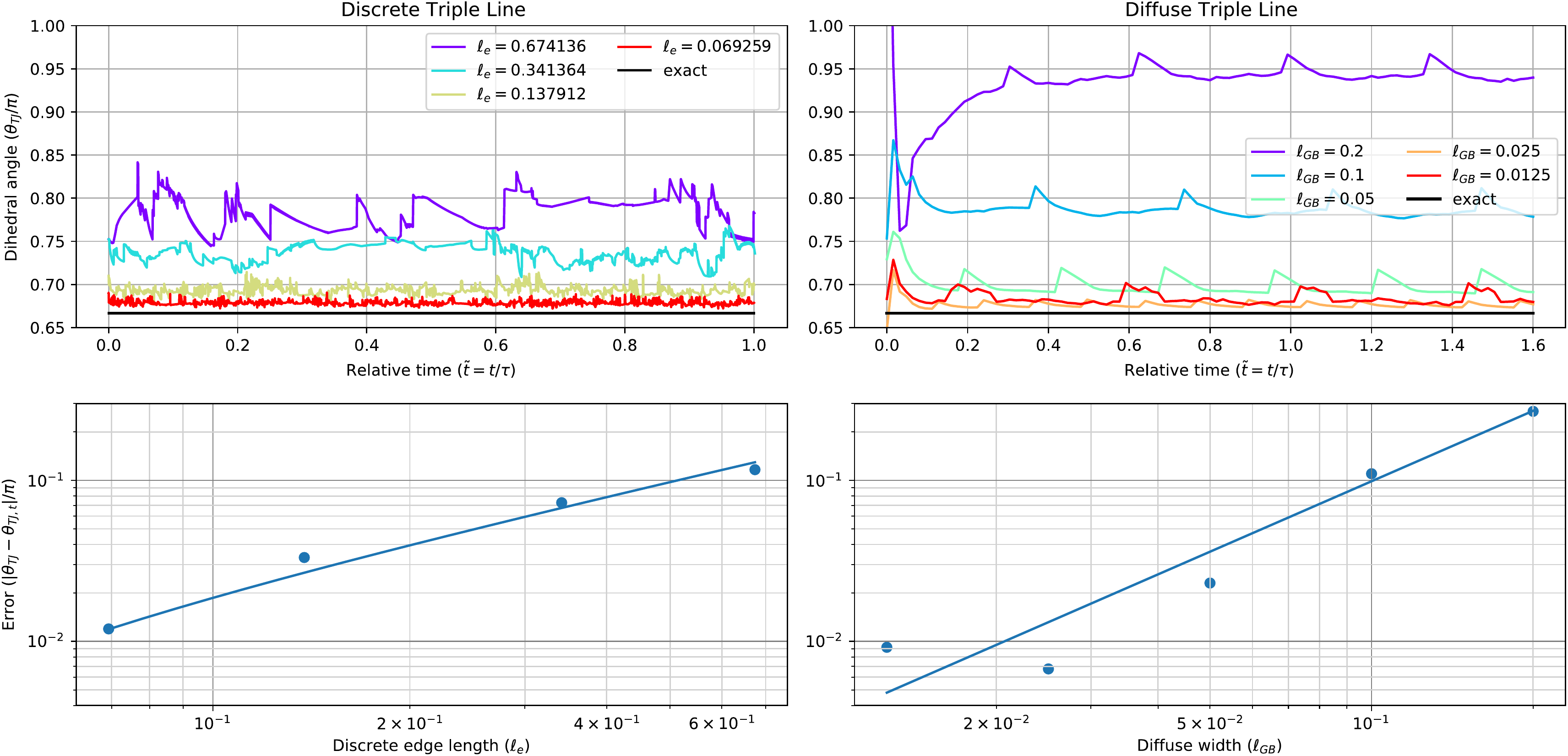}
  \caption{
    Comparison of $\theta_{TJ}$ for discrete model (left) and diffuse model (right); all quantities are nondimensionalized.
    (Top row) Plot of $\theta_{TJ}$ vs time, with color indicating the length scale and the exact solution in black.
    (Bottom row) Plot of the relative error vs length scale.} \label{fig:trip_ang11}
\end{figure}

The dihedral angle $\theta_{TJ}$ between the two boundaries of the blue grain is perhaps the simplest way to evaluate the accuracy of the geometry of the moving boundary in the vicinity of the TJ.
A force balance argument for constant and isotropic grain boundary properties (and in the absence of any TJ drag) leads to the condition $\theta_{TJ,t} = 2 \pi / 3$.
Moelans et al.\ \cite{moelans2009comparative} provides equations for the expected rate of area change and equilibrium junction angle for the more general situation where the grain boundary energies depend on misorientation, and uses these to evaluate the relative accuracy of two different diffuse interface methods, but does not perform a scaling analysis as is done below.
The expected equilibrium junction angle for the constant and isotropic grain boundary case is roughly enforced in the initial conditions by defining the two parts of the moving boundary to be the appropriate sections of of cylinders;
while this is not the steady-state profile given by Mullins, it is sufficiently close for a short initial transient and rapid convergence to the steady-state condition as is visible in \cref{fig:trip_ang11}.

As before, results for the discrete interface model are on the left and those for the diffuse interface model are on the right.
The top row shows $\theta_{TJ}$ as a function of time, where the color indicates the internal length scale and the exact solution $2 \pi / 3$ is in black.
The roughness of the curves for the discrete interface model is due to the remeshing required to maintain element quality, and the periodic spikes that appear for the diffuse interface model are due to the interaction of the adaptive mesh refinement and the construction of the isocontours.
The error in $\theta_{TJ}$ (measured as the median of the second half of the time series) is shown in the bottom row, with the dependence of the steady-state angle on $\ell_e$ for the discrete interface model being a consequence of the linear elements forcing the grain boundary curvature to be concentrated at the vertices and edges of the mesh.
Specifically, the grain boundary curvature that is distributed to the TJ edges causes the deviation of $\theta_{TJ}$ from the expected value, with the magnitude of the deviation depending on the product of $\ell_e$ and the mean curvature of the adjoining grain boundary.
Identifying the precise location of the TJ and the value of $\theta_{TJ}$ is more difficult for the diffuse interface model since the grain boundary geometry is implicit.
The procedure followed here involves fitting third- and fourth-order polynomial approximations to each side of the isocontour where the order parameter for the top grain is $0.5$.
The triple point location in the plane is then defined to be the point of intersection of the polynomials, and $\theta_{TJ}$ is the angle between the tangent vectors at the point of intersection.
This process works well in the sharp interface limit, but is very sensitive to perturbations in the solution for larger $\ell_{GB}$ since there is substantially more error in the predicted location of the TJ with respect to the simulation size.
The occasional deviations that are observed in the steady-state correspond to BSAMR re-gridding events.

Fitting a power law in the internal length scale $\ell$ to $|\theta_{TJ} - \theta_{TJ, t}| / \pi$ gives an exponent of $0.91 \pm 0.20$ for the discrete interface model and $1.45 \pm 0.13$ for the diffuse interface model, where the values are the medians and the uncertainties are half the interquartile range.
The additive offset of $(-0.005 \pm 0.011)\pi$ to the expected value of $\theta_{TJ}$ for the discrete interface model is entirely consistent with the TJ angle converging to the equilibrium angle in the $\ell_e \rightarrow 0$ limit, though at a lower rate than the half-life error magnitude in \cref{fig:sphere_r}.
This is not unexpected though, since the TJ can be thought of as a jump condition in the tangent plane to the grain boundary that is both difficult to accurately reproduce with a finite element mesh and is not present in the spherical grain case.
While the exponent for the diffuse case is nominally higher, this is not reflective of the trend observed for small $\ell_{GB}$ where the saturation in the error is likely the result of inaccuracy in the postprocess calculation of the angle.
The higher exponent therefore does not necessarily indicate better convergence.

\subsection{Quadruple point}
\label{subsec:quadruple_point}

\begin{figure}
\center
\subfloat[]{%
	\label{subfig1:hex}{%
		\includegraphics[height=30mm]{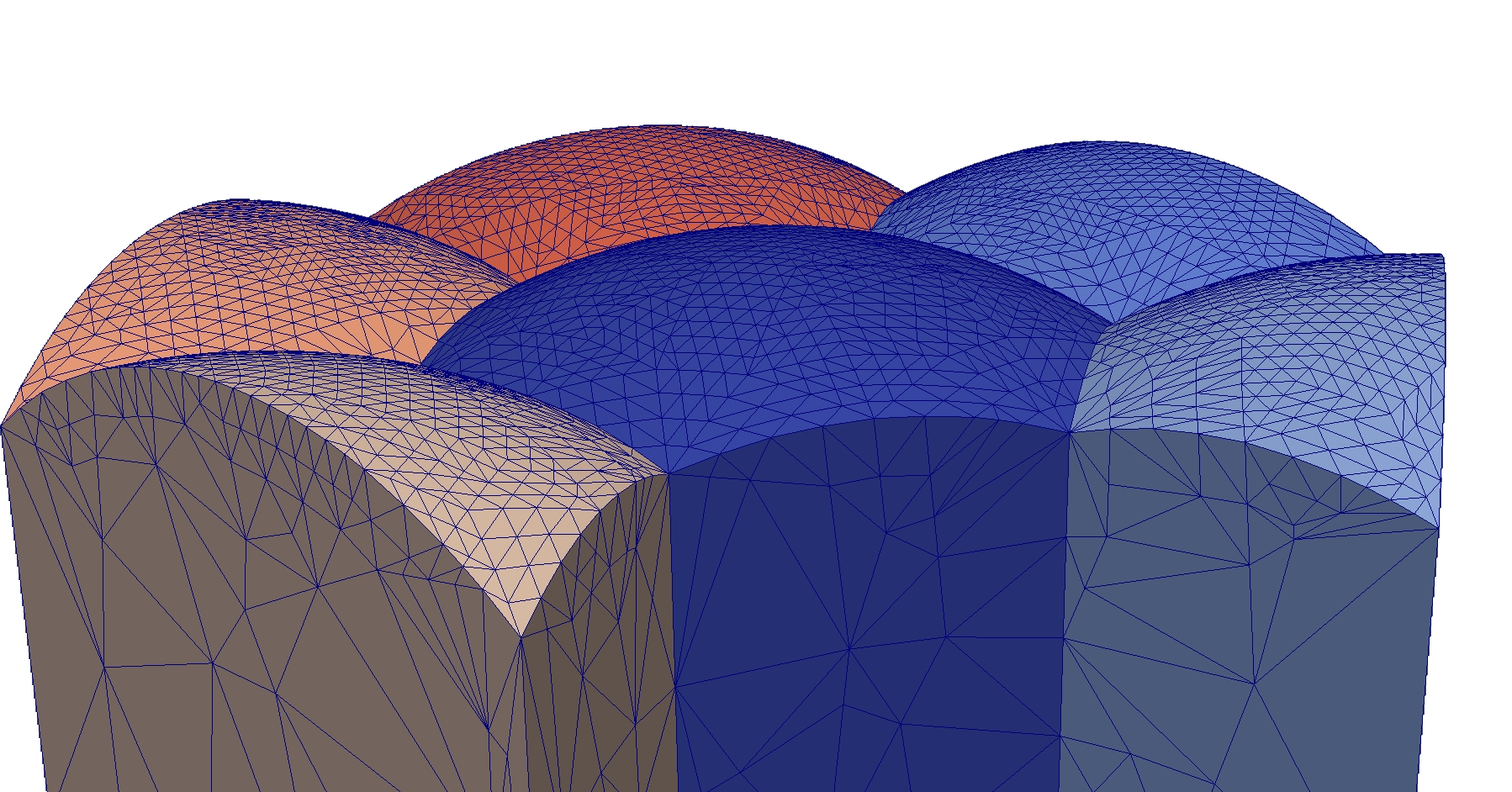}}}
\hspace{0em}
\subfloat[]{%
	\label{subfig1:hex_PF}{%
		\includegraphics[height=30mm]{%
			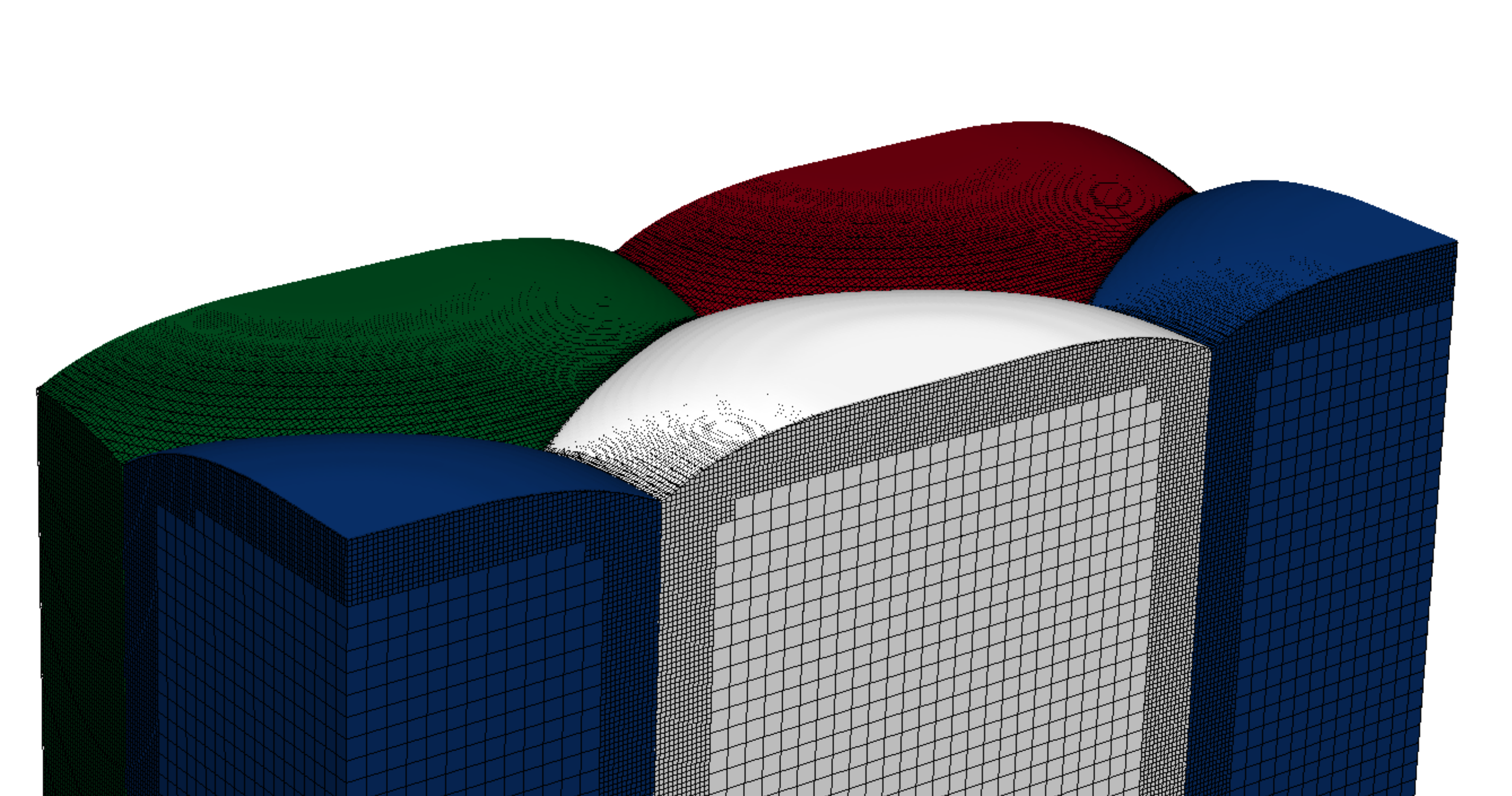}}}
\hspace{0em}
\subfloat[]{%
	\label{fig:quadruple_schematic}{%
		\includegraphics[height=30mm]{%
			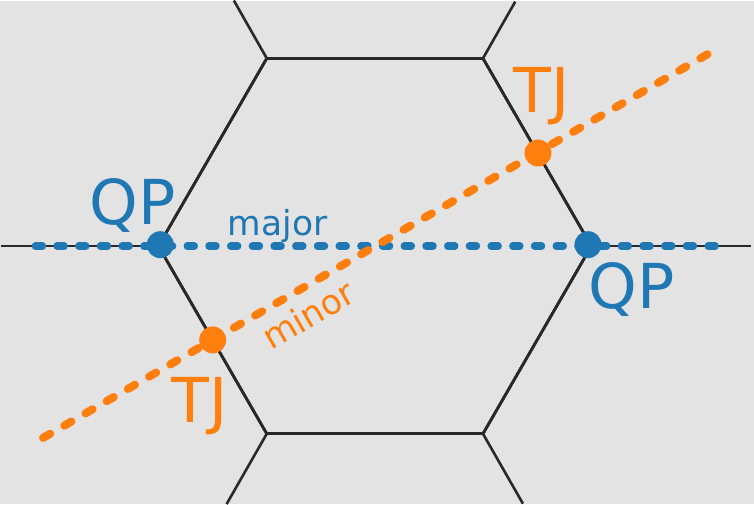}}}
\caption{\label{fig:quadruple}QP mesh configurations and schematic. (a) Hexagonal columnar grain mesh for the discrete interface model. (b) Hexagonal columnar grain BSAMR mesh in a rectilinear domain for the diffuse interface model. (c) Locations of QP and TJ along major and minor lines.}
\end{figure}

As with the TJ case, the grain structure for the QP case consists of a top grain above several columnar grains.
The grain boundaries of the top grain migrate down the simulation cell, consuming the columnar grains and eventually reaching a steady-state profile, though an analytical solution for this profile is not known.
The configurations of columnar grains for the discrete and diffuse interface models are shown in \cref{subfig1:hex,subfig1:hex_PF} respectively, with the hexagonal cross-sections of the columnar grains clearest for the discrete interface model;
the BSAMR mesh makes simulations of rectilinear domains like the one in the figure strongly preferable for the diffuse interface model.

\begin{figure}
  \includegraphics[width=\linewidth]{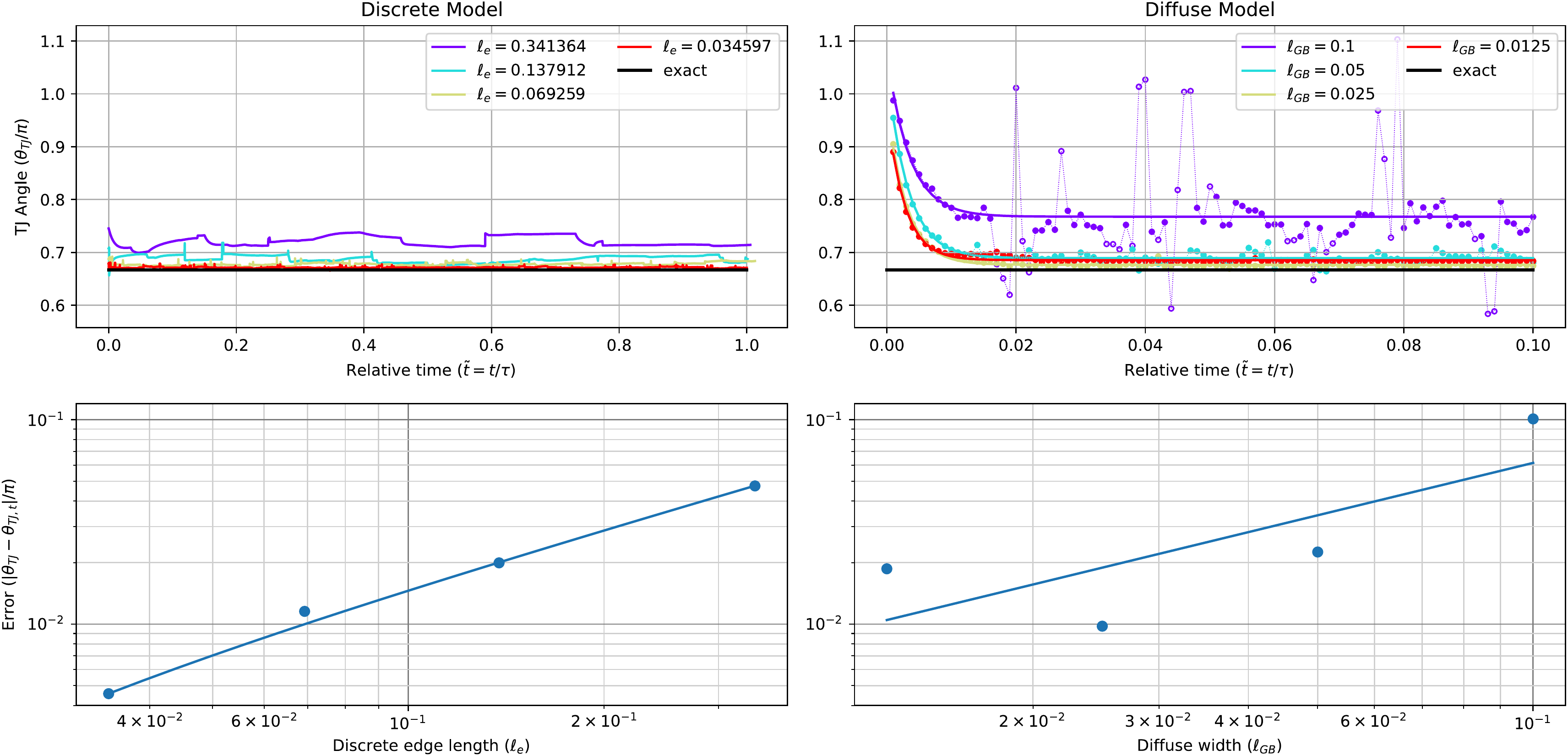}
  \caption{
    Comparison of minor axis results for the QP case for the discrete model (left) and diffuse model (right); all quantities are nondimensionalized.
    (Top row) Plot of the measured TJ (minor diameter) angle, with color indicating the length scale and the exact solution in black.
    (Bottom row) Plot of the relative error in the TJ angle with respect to length scale.
  } \label{fig:trip_ang12}
\end{figure}

\begin{figure}
  \includegraphics[width=\linewidth]{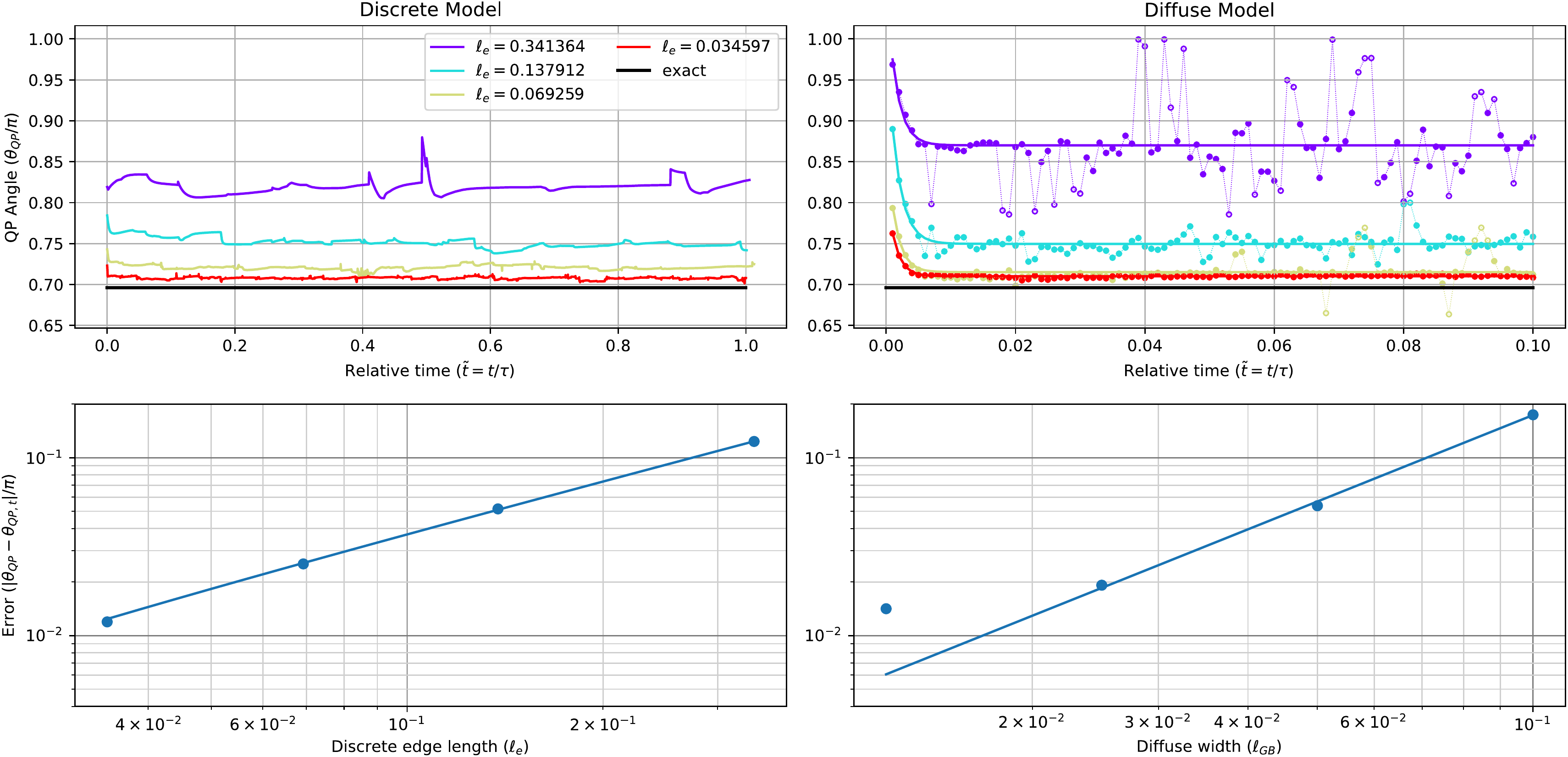}
  \caption{
    Comparison of major axis results for the QP case for the discrete model (left) and diffuse model (right); all quantities are nondimensionalized.
    (Top row) Plot of the measured QP (major diameter) angle, with color indicating the length scale and the exact solution in black.
    (Bottom row) Plot of the relative error in the QP angle with respect to length scale.
    %{\color{red} This figure is not referenced anywhere.}
  } \label{fig:trip_ang13}
\end{figure}

Following the initial transient, the steady-state profile is examined on the two planes indicated in \cref{fig:quadruple_schematic}, one along a minor diameter of the central grain and bisecting a TJ, the other along a major diameter of the central hexagonal grain and containing a QP.
The angles along these profiles at the intersections with the TJ and the QP are reported in \cref{fig:trip_ang12} and \cref{fig:trip_ang13}.
While the equilibrium angle at the TJ should be $2\pi/3$ (the same as for the TJ in \cref{subsec:triple_junction}), the curvature of the grain boundaries in both principal directions could change the rate of convergence to $2\pi/3$ with decreasing $\ell$ compared to the TJ case.
As for the equilibrium angle at the QP, an infinitesimal neighborhood of the QP will contain triple junction lines in a tetrahedral configuration connected by flat grain boundary surfaces provided the principal curvatures of the grain boundaries are finite.
This allows the equilibrium angle of $\cos^{-1}(-1/\sqrt{3}) \approx 0.696 \pi$ at the QP along the major diameter to be found by geometrical considerations.

%$a = 0.666 \pm 0.005$, $b = 0.131 \pm 0.018$, $c = 0.927 \pm 0.223$
Starting with the TJ angle, observe that the data points for the TJ angle error along the minor axis in the bottom row of Fig.\ \ref{fig:trip_ang12} closely resemble those for the TJ angle error in the bottom row of Fig.\ \ref{fig:trip_ang11}.
This indicates that the nonzero second principal curvature of the grain boundaries along the TJ lines in the QP case does not have a significant effect on the error in the equations of motion, and is consistent with the expectation that the error should scale with the mean curvature (the sum of the principal curvatures).
Fitting a power law in the internal length scale $\ell$ to $|\theta_{TJ} - \theta_{TJ, t}| / \pi$ gives an exponent of $0.927 \pm 0.223$ for the discrete interface model and $0.85 \pm 0.50$ for the diffuse interface model, with both models converging to the expected value.
While the exponent for the discrete interface model is nearly identical to that for the TJ case, the lower exponent for the diffuse interface model is likely a consequence of a power law fitting the data relatively poorly;
observe that the TJ angle error for the diffuse interface model does not fall on a line on a log-log plot, and instead seems to saturate at a lower bound set by the angle estimation procedure in postprocessing.

For the QP angle, the final values for the discrete interface model follow a power law in $\ell$ that converges to an angle of $(0.694 \pm 0.001) \pi$ with an exponent of $0.958 \pm 0.026$, whereas the respective values for the diffuse interface model are $(0.707 \pm 0.011) \pi$ and $0.85 \pm 0.45$; %$1.96651 \pm 1.070392$;
the limiting values for both the discrete and diffuse interface models effectively coincide with the exact value.
It is significant that the errors for all of the discrete interface results in Secs.\ \ref{subsec:triple_junction} and \ref{subsec:quadruple_point} decay with exponents that are close to one.
The discrete interface method uses linear elements that approximate the grain boundary geometry with first-order accuracy, meaning that an exponent of one is the best possible result.
It is likely that higher-order elements would need to be used to substantially increase the rate of error decay with $\ell_e$.
% Based on the geometric simplifications of Ref.\ \cite{2017ActMateMason}, the scaling of the errors for the TL and QP angles are estimated in Sec. \ref{app:error} to scale linearly with $\ell_e$ which match the observed scaling.
% Despite the low degrees of freedom allowed by the two triangles in the discrete interface method, both angle conditions converged to similar values for both methods. 
%Contrary to the findings of Ref.\ \cite{2017ActMateMason}, the results of \cref{subsec:spherical_grain} and this section suggest that the accuracy of the equations of motion depends on $\ell_e$ with an exponent of ${\sim}1.3$ independent of the vertex surroundings.
%The discrepancy with the literature could indicate a fortuitous cancellation of error in the rate of volume change for the highly symmetric configurations considered in Ref.\ \cite{2017ActMateMason}, and that does not occur in realistic conditions.
%This would make the exponent reported for the TJ case in \cref{subsec:triple_junction} the outlier, a reasonable possibility given the number of independent values of $\ell_e$ considered there.
% The diffuse model along the major axis (QP, \cref{fig:trip_ang12}), results in obtained parameters of $a = 0.70740 \pm 0.011076$, $b = 15.06544 \pm 36.992410$, and $c = 1.96651 \pm 1.070392$, which is similar to what the results of the discrete interface simulations indicate; the fitted exponent was $0.85 \pm 0.45$.
% Results for the diffuse boundary model indicate convergence to expected values for both the TJ and QP cases (\cref{fig:trip_ang12}).
The irregularity in the exponents for the diffuse interface model in Secs.\ \ref{subsec:triple_junction} and \ref{subsec:quadruple_point} is attributed to the error in the polynomial algorithm used to extract the grain boundary profile.
Examination of Figs.\ \ref{fig:trip_ang11} and \ref{fig:trip_ang12} indicates that this functions as a source of random error that is larger for highly diffuse boundaries but vanishes in the sharp boundary limit.

\section{Performance}
\label{sec:performance}

When selecting a numerical method in practice, computational cost is often nearly as much a concern as the accuracy of the simulated behavior.
This section specifically considers the dependence of the discrete interface method's computational cost on the internal length scale $\ell_e$;
given that the diffuse interface method's implementation \cite{zhang2019amrex,runnels2021massively} is considerably more mature than that of the discrete interface method \cite{2021PRMEren,2021VDlib}, and our concern is with asymptotic behavior rather than implementation specifics, a comparison with the diffuse interface method is omitted.
% The scaling can be estimated based on simplifying assumptions about the computational grid around the grain boundary interface and the mesh geometry for the diffuse and discrete interface methods, respectively.
% It is expected that the number of grid points in the diffuse interface method scales as $f(L/\ell_{GB})^3$ where $L$ is the characteristic length scale, $\ell_{GB}$ is the diffuse boundary width, and $f$ is the ratio of the grid height perpendicular to the grain boundary to the lateral dimensions of the grid. 
Suppose that the main contribution to the computational cost is evaluating the equations of motion for the grain boundary vertices.
The number of such vertices is expected to depend on the internal length scale as $\ell_{e}^{-2}$.
If the velocity of the vertices is independent of $\ell_{e}$, then the time step length should decrease as $\ell_{e}$ to keep the vertex displacement shorter than the characteristic edge length and prevent mesh element inversion.
This would imply that the overall computational cost should scale with $\ell_{e}^{-3}$, or as the product of the number of grain boundary mesh vertices and the number of time steps for a given overall simulation time.
% With increasing refinement, the time step should also decrease as $1/\ell$ for both methods to prevent element inversion for the discrete case, and numerical instability due to Courant–Friedrichs–Lewy (CFL) violation for the diffuse case.
% By this estimation, it is expected that the discrete interface method will outperform diffuse interface method as the refinement increases.

\begin{figure}
  \centering
    \includegraphics[width=0.5\linewidth]{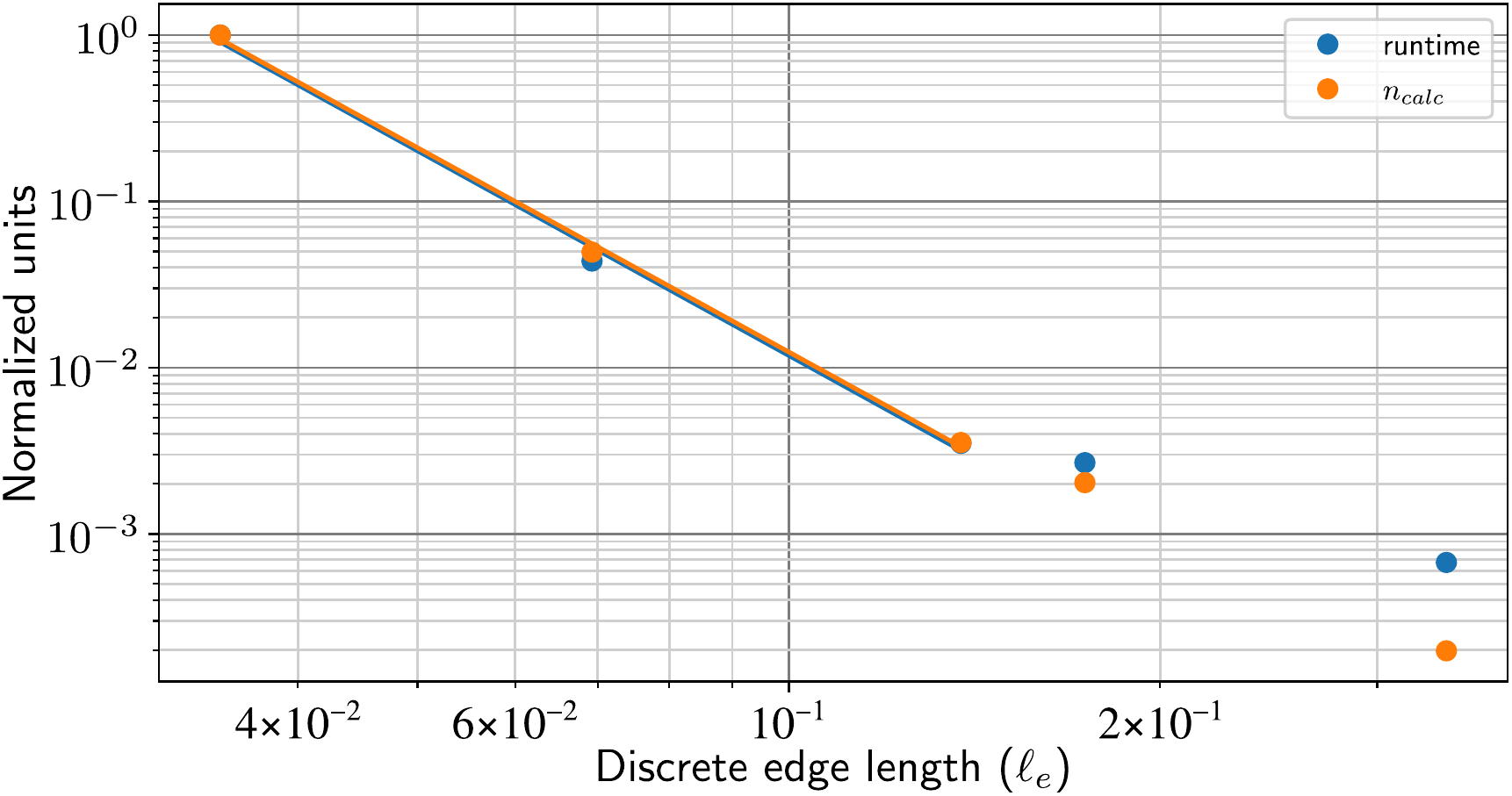}
    \caption{\label{fig:cost_discrete}The scaling of the normalized runtime and the normalized number of grain boundary vertex calculations for the spherical grain case as a function of $\ell_e$.}
\end{figure}

Figure \ref{fig:cost_discrete} shows the scaling of the normalized runtime cost and normalized number of grain boundary vertex calculations $n_{calc} = \sum_{j} n_{v_b, j}$, where $n_{v_b, j}$ is the number of grain boundary vertices at time step $j$, for the discrete interface method. 
These scale as $\ell_e^{-4.088}$ and $\ell_e^{-4.081}$, respectively, for small $\ell_e$ where the computational cost of the vertex calculations is expected to dominate. 
This confirms that the overhead of the discrete interface method (mesh management, enumeration of topological transitions, etc.) is relatively small compared to the evaluation of the equations of motion.
This overhead includes the local remeshing operations that are used to maintain the mesh quality and that occur at a frequency proportional to the time required for the interface to travel a distance $\ell_{e}$.
Further evidence that the computational cost of the remeshing operations is small relative to that of evaluating the equations of motion is given in Ref.\ \cite{2021PRMEren}, which also reports results for the evolution of a more extensive grain structure.
The scaling of the normalized runtime cost observed here is not consistent with the $\ell_e^{-3}$ scaling expected in the previous paragraph though.
This discrepancy is a result of the length of the median time step scaling as $\ell_e^{1.916}$ instead of linearly;
the underlying cause for this time step scaling is investigated further in \ref{app:performance}.

\section{Conclusion}
\label{sec:conclusion}

The purpose of this work has been to establish the validity and performance of a recently-developed discrete interface method by comparison to analytic solutions and a well-established multiphase field method.
More specifically, the evolution of the simplest configurations involving surfaces, triple lines, and quadruple points with self-similar behavior given constant and isotropic grain boundary properties are used to quantify the error in position and junction angles as a function of the degree of refinement.
The boundary types are simple enough to be amenable to analysis, yet complex enough to introduce different systematic errors over the course of their evolution.
Despite the approaches for simulating boundary motion being distinctly different, our results indicate that both methods converge to the same junction angles with similar rates.
The most significant difference is that when predicting the half life of the shrinking sphere, the convergence rate of the diffuse interface method appears to be about half of that of the discrete interface method.
% (The discrete interface method has a slightly higher convergence rate in prediction of the sphere half-life as an indicator for estimation of the time for topological transitions.)
%Using discrete interface method with linear elements, an important mesh consideration becomes apparent when investigating the angle of the boundary trace along the major axis plane of the quadruple angle case. 
% The diffuse interface method has a clearly convergent behavior of the junction angle, yet in the discrete interface method, due to lower degrees of freedom caused by representing the surface corner by two triangles for each case, the junction angle converges to a different value with a slower convergence rate.
% Although the overall evolution of the boundary is similar, this indicates the importance of meshing in representing the grain boundary orientation which is expected to become important when simulating under anisotropic grain boundary properties.
% Both methods show a clearly convergent behavior to the exact angle conditions, except for the trace of the angle along the major axis plane which has a higher uncertainty in the convergence rate.

Although this work assumes constant and isotropic grain boundary properties, both methods were developed with the intention of performing simulations for anisotropic grain boundary properties.
The integration of an accurate grain boundary energy for arbitrary orientation relationships and interface orientations is a current challenge in microstructure modeling. 
Morawiec \cite{2000ActaMateMorawiec} suggested that the grain boundary energy could be experimentally obtained as a function of the grain boundary crystallography by applying the Herring condition \cite{1951herring,1953herring} to triple junctions imaged by three-dimensional microscopy techniques \cite{2005MatSciFormZaefferer, 2013ApplCrysLi}.
Alternatively, molecular dynamics simulations allow direct evaluation of grain boundary properties in bicrystals for a large but not exhaustive subset of the five-dimensional grain boundary space \cite{2009ActaMateOlmsted_I} .
% Both approaches are subject to the excessive number of sampled points required to adequately characterize the grain boundary energy as a function on a five-dimensional space, estimated to be $2 \times 10^5 $ for $5^\circ$ angular resolution \cite{2000IntSciSaylor} and $6.5 \times 10^8$ for $1^\circ$ angular resolution for cubic systems;
% the grain boundary mobility is subject to analogous requirements.
While the excessive number of points required to adequately sample this space of has precluded the availability of general grain boundary energy and mobility functions in the literature, there has been progress for particular subsets of grain boundaries \cite{2014ActMateBulatov}.
Models have also been presented that can accurately predict grain boundary energy for most known orientations
\cite{2016JourMechPhySolidRunnels}, and have been used in combination with phase field to predict such behavior as faceting and disconnection migration \cite{gokuli2021multiphase}.
However, in addition to the general problem of obtaining accurate grain boundary energy, the nonconvexity of this function induces numerical issues that must be handled explicitly \cite{gokuli2021multiphase}. The extension of this present framework in that direction shall therefore constitute future work.

The performance of the discrete interface model lends confidence in its ability to yield accurate results for more general and complex microstructures for which there is no known analytic solution.
Moreover, the performance with unoptimized code indicates reasonable scaling behavior that is close to the ideal scaling and comparable to that of alternative methods.

\section*{Acknowledgements}
\label{sec:ack}

EE and JKM were supported by the National Science Foundation under Grant No.\ DMR 1839370.
BR was supported by the Office of Naval Research, grant \#N00014-21-1-2113.
This work used the INCLINE cluster at the University of Colorado Colorado Springs.
INCLINE is supported by the National Science Foundation, grant \#2017917.

\section{Data availability}
The Alamo (\url{https://github.com/solidsgroup/alamo}) and VDlib (\url{https://github.com/erdemeren/VDlib}) libraries used to generate these results are available as open source. The processed data required to reproduce these findings are available to download from [\url{https://arxiv.org/abs/2203.03167}].

\FloatBarrier

\bibliographystyle{ieeetr} 
\bibliography{main}

\appendix

\section{Nondimensionalization}
\label{app:nondim}

Define the variable $L$ to be the characteristic length scale of the grain structure defined in \cref{sec:results}.
For a sphere it is the sphere radius, for the TJ it is the length of the simulation cell in the direction normal to the consumed grain boundary, and for the QP it is the hexagonal grains's minor diameter.
The Turnbull equation in \cref{eqn:Turnbull} suggests that there is a characteristic time scale $\tau = L^2 / (m \gamma)$.
The simulations are performed with nondimensionalized time $\tilde{t} = t / \tau$, nondimensionalized space $\tilde{x} = x / L$, nondimensionalized rate of volume change $d\tilde{V} / d\tilde{t} = (\tau / L^3) dV / dt$, etc.

With respect to the quantities defined in \cref{subsec:discrete_interface}, suppose that $\tau_l(\nvec{t}_i) = 0$, $\delta_0 = 0$, $\delta_1(\nvec{t}_i) = 0$, and that $\gamma(\nvec{n}_{ij})$ and $\delta_2(\nvec{n}_{ij})$ are constants.
The governing equations of the discrete interface model then reduce to:
\begin{align}
  \bm{F} &= \frac{\gamma}{2} \sum_{i} ||\bm{t}_i|| \sum_{j:\{i,j\}\in \Delta} \hat{\bm{n}}_{ij} \times \hat{\bm{t}}_i,\\
  \bm{D} &= \frac{\delta_2}{6} \sum_{i, j \in \Delta}
    || \bm{t}_i \times \bm{t}_j || (\nvec{n}_{ij} \otimes \nvec{n}_{ij}),\\
  \bm{v} &= \bm{D}^{-1} \bm{F}.
\end{align}
The nondimensionalized versions of these equations are
\begin{align}
  &\tilde{\bm{F}} = \frac{\bm{F}}{L\gamma}  = \frac{1}{2} \sum_{i} ||\tilde{\bm{t}_i}|| \sum_{j:\{i,j\}\in \Delta} \hat{\bm{n}}_{ij} \times \hat{\bm{t}}_i,\\
  &\tilde{\bm{D}} = \frac{\bm{D}}{L^2 \delta_2} = 
    \frac{1}{6} \sum_{i, j \in \Delta} 
    ||\tilde{\bm{t}}_i \times \tilde{\bm{t}}_j || (\hat{\bm{n}}_{ij} \otimes \hat{\bm{n}}_{ij}),\\
  &\tilde{\bm{v}} = \frac{\tau}{L}\bm{v} = \frac{\tau \gamma}{L^2 \delta_2} \tilde{\bm{D}}^{-1} \tilde{\bm{F}},
\end{align}
where $\delta_2 = 3 / m$ when the triple line and quadruple point drags vanish;
this can be derived by requiring that the limiting behavior of a small spherical cap coincides with the predictions of \cref{eqn:Turnbull}.

The corresponding nondimensionalization of the multiphase field governing equations in Sec.\ \ref{subsec:diffuse_interface} yields
\begin{align}
    \frac{\partial \eta}{\partial \tilde{t}} &= -\tau L \frac{\delta W}{\delta \eta_n}
    = \frac{\partial \tilde{w}}{\partial \eta_n} + k\tilde{\Delta} \eta_n, &
    \tilde{w} &= w\,\tau\,L, &
    \tilde{\Delta} = \tau L k\Delta.
\end{align}
In this work, all multiphase field calculations are performed with dimensional values and then nondimensionalized for comparison to discrete interface simulations.

\section{Spherical grain}
\label{app:sphere}

If grain boundary properties are constant and isotropic, then the evolution of a spherical grain is self-similar and can be completely described by the radius $r_t(t)$ as a function of time.
Since the mean curvature is $K = 2 / r_t$, \cref{eqn:Turnbull} implies that
\begin{equation}
  dr_t / dt = -2 m \gamma / r_t.
\end{equation}
Setting $t_0 = 0$ and $r_t(0) = r_0$ and integrating gives
\begin{equation} 
  r_t(t) = \sqrt{r_0^2 - 4 m \gamma t}
\end{equation}
as the solution to this differential equation.
Since the characteristic length scale for a sphere is $r_0$, nondimensionalizing reduces this to
\begin{equation} 
  \tilde{r}_t(\tilde{t}) = \sqrt{1 - 4 \tilde{t}}
\end{equation}
for the black curve in \cref{fig:sphere_r}.

% \section{Angular error estimation} \label{app:error}

% \begin{figure}
%   \centering
%     \includegraphics[width=\linewidth]{./results_discrete/error/err_est.pdf}
%     \caption{\label{fig:err_est_TL_QP} The convergence of error in predicting (a) TL and (b) QP angles, under the idealized boundary assumptions.}
% \end{figure}

% The error in the rate of volume change is estimated in Ref. \cite{2017ActMateMason} by idealizing the boundaries around the triple junctions and quadruple points as portions of Reuleaux triangles and spherical tetrahedron, respectively. 
% Assuming that the discretized vertices lie on these idealized boundaries, the convergence of error for the angles can be estimated by varying $\ell_e$. 
% The junction configurations considered in this article are slightly different, but the idea is the same. 
% Figure \ref{fig:err_est_TL_QP} shows that for $\ell_e < 0.4$ the errors are estimated to scale as $\ell_e^{1.001}$ and $\ell_e^{0.990}$.

\section{Scaling analysis}
\label{app:performance}

As described in Sec.\ \ref{sec:performance}, while the computational cost of the discrete interface method is expected to scale as $\ell_e^{-3}$, the actual scaling is instead $\ell_e^{-4.089}$.
Closer investigation revealed that the time step could decrease or increase by multiple orders of magnitude depending on the presence of various local mesh configurations.
The boundary triangles exert capillary forces only in the boundary plane, yet contribute drag forces only in the out-of-plane direction.
This allows vertices on nearly-flat grain boundary sections to experience arbitrarily large lateral velocities, slowing the simulation down as the time step is reduced to prevent element inversion. 
% This can be most easily imagined by considering a closed disk of coplanar equilateral triangles.
% Initially the net capillary force will be zero, and the eigendirection of the drag tensor with a non-zero eigenvalue is in the plane normal direction.
%  However, if one moves the central vertex slightly in plane, the net force will be non-zero and $\bm{v} = \bm{D}^{-1} \bm{F} \rightarrow \infty$.
The discrete method simulations in Sec.\ \ref{subsec:spherical_grain} include an isotropic contribution $\bm{D}_{I, d} = A^2_{m}/(m d) \bm{I}$ to the drag tensor such that $\bm{v} = (\bm{D} + \bm{D}_{I, d})^{-1} \bm{F}$, where $A^2_{m}$ is the mean triangle area over the whole simulation, $m$ is the mobility, and $d = 1000$ is a drag ratio.
Decreasing the drag ratio reduces the lateral velocities, but also slows down the actual motion of the boundary and introduces a systematic error. 

\begin{figure}
  \centering
    \includegraphics[width=0.35\linewidth]{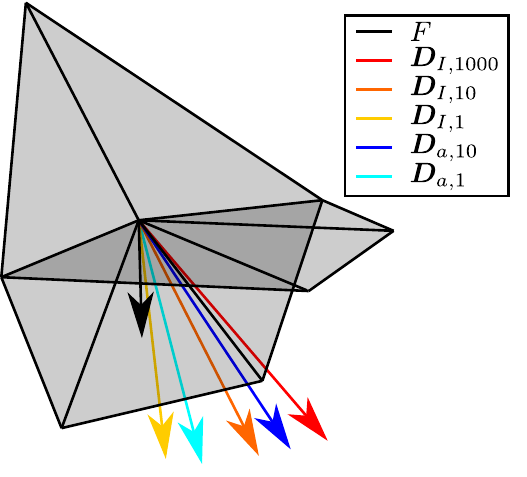}
    \caption{\label{subfig:drag_corr} The effect of different drag tensor correction terms on the resulting velocity. The capillary force is colored black and the velocities corresponding to different correction terms are differentiated by color. Each vector is scaled relative to the maximum magnitude among the velocities.}
    %\caption{\label{subfig:drag_corr} The effect of different drag tensor correction terms on the resulting velocity. The capillary force is colored black and the velocities corresponding to the $\bm{D}_{I, 1000}$, $\bm{D}_{I, 10}$, $\bm{D}_{I, 1}$, $\bm{D}_{a, 10}$, and $\bm{D}_{a, 1}$ are colored red, cyan, cyan, blue, and blue, respectively. Each vector is scaled relative to the maximum magnitude among the velocities.}
\end{figure}

As an alternative, a contribution to the drag tensor that only acts in the in-plane directions could be constructed as follows.
%For simplicity, consider a closed disk of coplanar triangles around a vertex, and the tensor formed by the sum of the outer products of the $2$-stratum edge directions for the edges in the triangles across the vertex. 
For simplicity, consider a closed disk of coplanar triangles around a vertex. 
Iterating over each grain boundary triangle $\Delta_{ij}$ adjacent to the central vertex, find the relative positions of the other vertices from the central vertex $\bm{p}_i$ and $\bm{p}_j$ and construct the outer product of the difference $\bm{p}_i - \bm{p}_j$ with itself.
Let $\Lambda_{max}$ be the largest eigenvalue of the sum of the outer products, and define the matrix $\bm{C} = \sum_{i,j \in \Delta} (\bm{p}_i - \bm{p}_j) \otimes (\bm{p}_i - \bm{p}_j) / \Lambda_{max}$.
%Iterating over each triangle, calculate the outer product of the displacement vector corresponding to the edge across the vertex with itself.
%Consider the tensor formed by summing these outer products, and denote by $\bm{C}$ this tensor divided by its largest eigenvalue.
The anisotropic drag tensor contribution $\bm{D}_{a, d} = A^2_{m}/(m d) \bm{C}$ by construction has no effect on the grain boundary motion in the plane normal direction.
This should allow the lateral velocities of boundary vertices to be reduced while introducing less systematic error in the motion of non-planar boundaries than for an isotropic drag.
An example triple junction mesh configuration is shown in \cref{subfig:drag_corr} to qualitatively demonstrate the effect of different drag tensor correction terms.
Although the velocity associated with $\bm{D}_{I, d}$ aligns with the force direction faster with increasing $d$, the velocity term in the vertical direction is also attenuated more compared to $\bm{D}_{a, d}$. 

\begin{figure}
  \centering
    \includegraphics[width=0.5\linewidth]{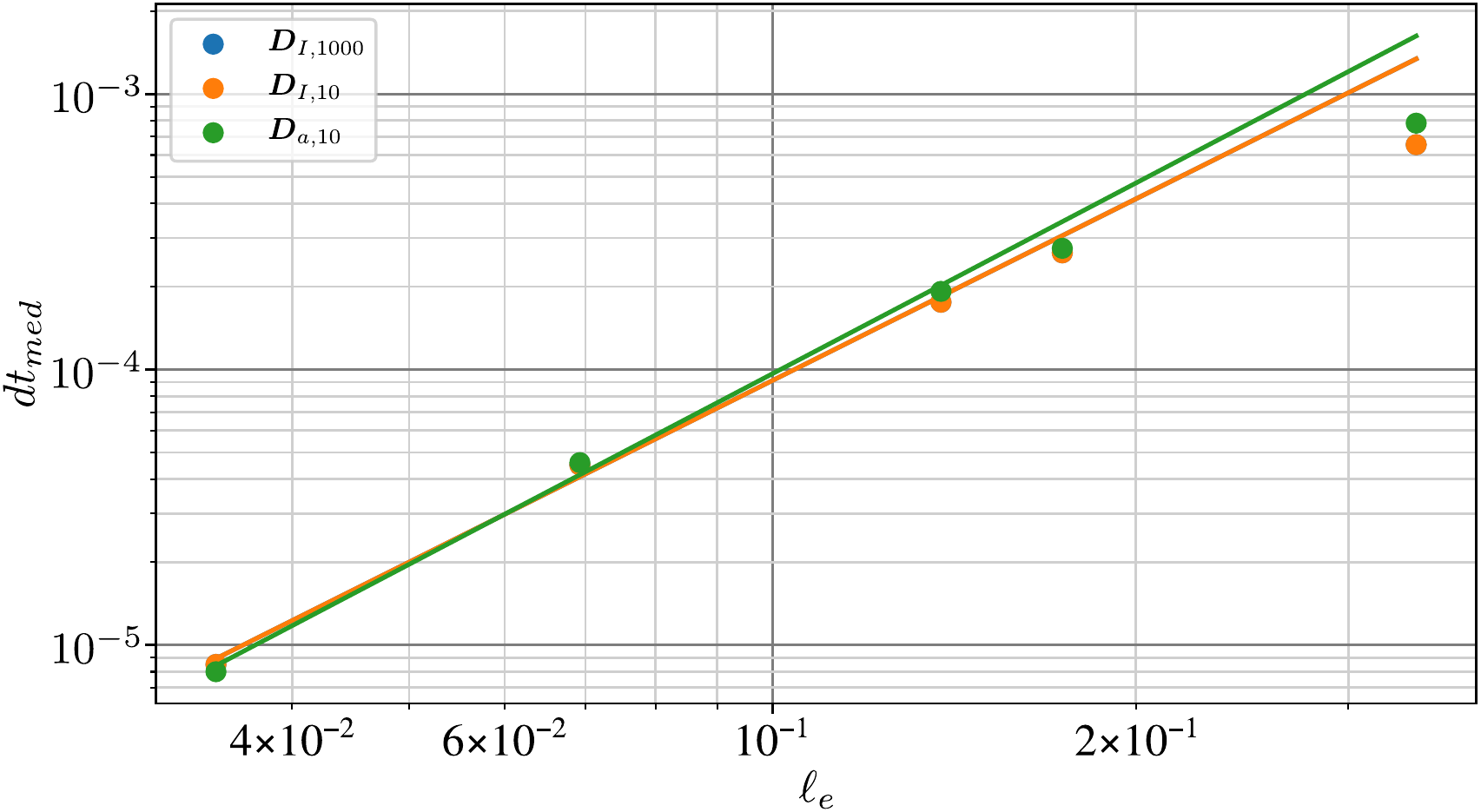}
    \caption{\label{fig:dt_vs_l} The scaling of the median time step with $\ell_e$ for the three drag tensor correction terms.}
\end{figure}

The difference in the expected and the actual scaling of the cost can largely be attributed to the non-linear scaling of the median time step $dt_{med}$ shown in \cref{fig:dt_vs_l}.
It scales as $\ell_e^{2.010}$ for $\bm{D}_{a, 10}$ and $\ell_e^{1.916}$ for $\bm{D}_{I, 1000}$ and $\bm{D}_{I, 10}$. 
Overall, $\bm{D}_{a, 10}$ allows larger time steps and has a better accuracy, though the improvement is not significant.

% \section{Triple junction}
% \label{app:triple}
% 
% For the triple junction case, After the simulation reaches steady state regime, the profiles for the grain boundaries on the cross-section plane given by $y = 0$ can be extracted as shown in Fig. \ref{app:quadruple} in comparison to the analytical solution. The error $|z_{rel}(x) - z_{rel,t}(x)|$ converges as ... where $z_{rel}(x) = z(x) - z_{tl}$.
% \todo{add the analytical profile, calculate the error in vertical displacement with respect to the triple junction position}
% \begin{figure}[H]
% \centering
% \begin{subfigure}{0.49\linewidth}
% \includegraphics[height=0.6\linewidth]{results_discrete/TripleJunction/slices_trip010.pdf}
% \caption{}
% \end{subfigure}
% \begin{subfigure}{0.49\linewidth}
% \includegraphics[height=0.6\linewidth]{results_discrete/TripleJunction/slices_trip010_coincide.pdf}
% \caption{}
% \end{subfigure}
% 
% \caption{The cross-sections of the moving boundaries of the triple junction, along the triple junction direction for different levels of refinement, displaced for easier identification.}
% \end{figure}
% \todo{Add the profile for the analytical case}

% \section{Quadruple point} \label{app:quadruple}

\end{document}